\documentclass{aa}  

\usepackage{graphicx}
\usepackage{longtable}
\usepackage{lscape}
\usepackage{natbib}
\usepackage{lscape}
\usepackage[varg]{txfonts}
\begin{document}
   \title{Binary frequency of planet-host stars at wide separations}

   \subtitle{A new brown dwarf companion to a planet-host star}

   \author{N. Lodieu \inst{1,2}\thanks{Based on observations collected at the
   European Organisation for Astronomical Research in the Southern Hemisphere, Chile.}
   \thanks{Based on observations made with the Gran Telescopio Canarias (GTC),
installed in the Spanish Observatorio del Roque de los Muchachos of the Instituto
de Astrof\'isica de Canarias, in the island of La Palma (program program GTC65-13B)}
          \and
          A. P\'erez-Garrido \inst{3}
          \and
          V.\ J.\ S.\ B\'ejar \inst{1,2}
          \and
          B.\ Gauza \inst{1}
          \and
          M. T. Ruiz \inst{4}
          \and
          R.\ Rebolo \inst{1,2,5}
          \and
          D.\ J.\ Pinfield \inst{6}
          \and
          E.\ L.\ Mart\'in \inst{7}
          }

   \institute{Instituto de Astrof\'isica de Canarias (IAC), Calle V\'ia L\'actea s/n, E-38200 La Laguna, Tenerife, Spain \\
         \email{nlodieu,vbejar,bgauza,rrl@iac.es}
         \and
         Departamento de Astrof\'isica, Universidad de La Laguna (ULL),
E-38206 La Laguna, Tenerife, Spain
         \and
         Universidad Polit\'ecnica de Cartagena, Campus Muralla del Mar, Cartagena, Murcia E-30202 Spain
         \and
         Departamento de Astronom\'ia, Universidad de Chile, Casilla 36-D, Santiago, Chile
         \and
         Consejo Superior de Investigaciones Cient\'ificas, CSIC, Spain
         \and
         Centre for Astrophysics Research, Science and Technology Research Institute, University of Hertfordshire, 
Hatfield AL10 9AB, UK
         \and
         Centro de Astrobiolog\'ia (INTA-CSIC), Departamento de Astrof\'isica. P.O. Box 78, E-28691 Villanueva de la Ca\~nada, Madrid, Spain
             }

   \date{\today{}; \today{}}

 
  \abstract
   {}
   {The aim of the project is to improve our knowledge on the multiplicity
of planet-host stars at wide physical separations.}
   {We cross-matched approximately 6200 square degree area of the Southern sky imaged by 
   the Visible Infrared Survey Telescope for Astronomy (VISTA) Hemisphere Survey (VHS) 
   with the Two Micron All Sky Survey (2MASS) to look for wide common proper motion companions
   to known planet-host stars. We complemented our astrometric search with photometric
   criteria.}
   {We confirmed spectroscopically the co-moving nature of seven sources out of
16 companion candidates and discarded eight, while the remaining one stays as
a candidate. Among these new wide companions to planet-host stars, we discovered
a T4.5 dwarf companion at 6.3 arcmin ($\sim$9000 au) from HIP\,70849, a K7V star 
which hosts a 9 Jupiter mass planet with an eccentric orbit. We also report two 
new stellar M dwarf companions to one G and one metal-rich K star. 
We infer stellar and substellar binary frequencies for our complete sample of 
37 targets of 5.4$\pm$3.8\% and 2.7$\pm$2.7\% (1$\sigma$ confidence level), 
respectively, for projected physical separations larger than 
$\sim$60--160 au assuming the range of distances of planet-host stars (24--75 pc).
These values are comparable to the frequencies of non planet-host stars.
We find that the period-eccentricity trend holds with a lack of multiple systems
with planets at large eccentricities (e\,$>$\,0.2) for periods less than 40 days. 
However, the lack of planets more massive than 2.5 Jupiter masses and short periods 
($<$40 days) orbiting single stars is not so obvious due to recent discoveries
by ground-based transit surveys and space missions.
}
   {}


   \maketitle
%

%
%
\section{Introduction}
\label{compHIP:intro}

The discoveries of the first extrasolar planet orbiting a solar-type star and the first 
brown dwarfs were announced the same year \citep{mayor95,rebolo95,nakajima95}.
Since then, more than one thousand exoplanets have been discovered, mainly thanks to radial
velocity surveys and transit campaigns but also through direct imaging, microlensing,
and timing\footnote{see the exoplanet.eu repositary for a complete list of planets
and references}. In the meantime, a similar number of nearby field L, T, and Y dwarfs 
($\sim$1300)\footnote{the full list is available in DwarfArchives.org}, have been 
announced thanks to the advent of large-scale surveys such as
the DEep Near Infrared Survey \citep[DENIS; e.g.][]{epchtein97,delfosse97,martin97b,martin99a}, 
the Two Micron All Sky Survey \citep[2MASS; e.g.][]{skrutskie06,burgasser99,burgasser03c,cruz07}, 
the Sloan Digital Sky Survey \citep[SDSS; e.g.][]{york00,leggett00,geballe02,chiu06}, 
the UKIRT Infrared Deep Sky Survey \citep[UKIDSS;][]{lawrence07,lodieu07b,pinfield08,burningham10b,burningham13}, 
and the Wide-field Infrared Survey Explorer \citep[WISE;][]{wright10,kirkpatrick11,mace13,thompson13}. 
However, only one wide brown dwarf companion co-moving with a planet-host star has been reported
to date, HD\,3651B \citep[T7.5; d\,=\,11.1$\pm$0.1 pc;][]{mugrauer06b,liu07,leggett07b,burgasser07d}.

The presence of companion(s) to planet-host stars is thought to affect the efficiency
of planet formation as well as the properties of any planet orbiting the primary 
star \citep{boss06a,nelson03a,moriwaki04} due to several processes, 
such as gravitational instability, accretion, 
velocities of colliding planetesimals, dissipation, and/or Kozai effects.
Looking at a sample of close and wide companions to planet-host stars can
reveal peculiarities in the eccentricity or mass distribution within those
systems that may constrain current models of planet formation. In other
words, one would like to address the following question: do companions to 
planet-host stars influence the properties of the planetary systems and to 
what degree?

Some planet-host stars are members of well-known multiple systems, such as
16\,Cyg for example \citep{mason01}. \citet{eggenberger04} announced a sample 
of 15 new wide low-mass companions to planet-host stars and suggested a possible
difference in the mass-period and eccentricity-period distributions, trend
previously reported in \citet{zucker02}. These discoveries and potential
differences in the properties of planets orbiting single stars and planets
orbiting stars in multiple systems triggered dedicated searches by several groups
\citep{butler97,mugrauer04a,mugrauer04b,konacki05,mugrauer05b,mugrauer05c,eggenberger06,mugrauer06a,mugrauer07a,mugrauer07b,desidera07,bonavita07}
to confirm (or otherwise) those trends in the properties of exoplanets.
These searches also led to the discoveries of the first wide brown dwarf companion
to HD\,3651 \citep{mugrauer06b}, a K7 dwarf with a sub-Saturn mass planet 
\citep{fischer04a,santos05a}, and the most evolved multiple system with an exoplanet
\citep{butler01}, composed of a K subgiant and a white dwarf 
\citep[HD\,27442;][]{chauvin06,raghavan06,mugrauer07a}.
Moreover, several close companions have been announced around planet-host
stars, thanks to high-resolution adaptive optics imaging \citep{els01,patience02}
and lucky imaging \citep{ginski12}.
All these surveys point towards a global multiplicity of at least 20\% among solar-type
stars with exoplanets \citep{mugrauer07a}.

In this paper, we describe a new search for wide companions to planet-host stars
taking advantage of the homogeneous coverage of the Visible and Infrared Survey 
Telescope for Astronomy \citep[VISTA;][]{emerson01,emerson04,dalton06} Hemisphere
Survey \citep[VHS;][]{mcmahon12} that will provide a full coverage of the Southern 
sky in the upcoming years. We cross-correlated VHS with 2MASS \citep{cutri03,skrutskie06}
over an area of $\sim$6200 square degrees.
Our work represents an extension and a complementary approach to the 
techniques employed by other authors \citep[e.g.][]{mugrauer07a}.
In Section \ref{compHIP:surveys} we present the characteristics of the 2MASS 
and VHS surveys as well as our astrometric and photometric search criteria.
In Section \ref{compHIP:spectro} we describe our optical and near-infrared
spectroscopic follow-up of companion candidates conducted with the spectrographs 
mounted on the New Technology Telescope (NTT) at La Silla Observatory in Chile
and the OSIRIS instrument on the 10.4-m Gran Telescopio de Canarias at the 
Roque de Los Muchachos observatory on the island of La Palma (Canary Islands, Spain).
In Sections \ref{compHIP:optical_SpT} and \ref{compHIP:AbsMag_SpT} we present
the optical spectral classification scheme and the absolute magnitude vs
spectral type relation employed throughout our analysis.
In Section \ref{compHIP:new_comp} we analyse the properties of the
new wide companions found in our sample, including a brown dwarf binary
candidate co-moving with HIP\,70849\@.
In Sections \ref{compHIP:stability}--\ref{compHIP:properties} we discuss the 
stability of the
widest companions in our sample, derive the frequency of planet-host stars
at wide separations, and examine the dynamical properties of the systems
with respect to previous studies \citep{eggenberger04,mugrauer07a,desidera07}.

%
%
\section{Surveys and search criteria}
\label{compHIP:surveys}
\subsection{The 2MASS database}
\label{compHIP:surveys_2mass}

The 2MASS project imaged the full sky in three near-infrared filters, 
$J$ (1.25 $\mu$m), $H$ (1.65 $\mu$m), and $K_{s}$ (2.17 $\mu$m), down to 
10$\sigma$ depths of 15.8 mag, 15.1 mag, and 14.3 mag, 
respectively \citep{skrutskie06}\footnote{http://www.ipac.caltech.edu/2mass/}. 
The survey was conducted with 256$\times$256 HgCdTe detectors installed on two 
automated 1.3-m telescopes located on Mount Hopkins in Arizona (USA) and on
Cerro Tololo (CTIO) in Chile. The observing strategy consisted of scanning the
sky in three filters simultaneously with patches on the sky covering six degrees
in declination and 8.5 arcmin in right ascension. The Northern site started 
operations in 1997, followed by the Southern site a year later. The survey 
ended at the beginning of 2001, providing a final world-public catalogue of 
$\sim$471 million sources \citep{cutri03}. Comparison of the 2MASS catalogue 
with other catalogues such as Tycho-2 \citep{hog00} and UCAC \citep{zacharias03}
indicate that 2MASS positions in the International Celestial Reference 
System are accurate to 70-80 mas (root-mean-square; rms) over most of the magnitude 
range ($K_{s}$ between 6 and 15 mag). This all-sky survey played a major 
role in astronomy over the past decade, in particular in our knowledge of the 
low-mass and substellar populations of the solar neighbourhood 
\cite[e.g.][]{reid99a,burgasser99,burgasser04d,cruz07,kirkpatrick10}.

\subsection{The VISTA Hemisphere survey}
\label{compHIP:surveys_VHS}

VISTA is a 4-m telescope \citep{emerson01,emerson04} located at the Cerro
Paranal Observatory (Chile) run by the  European Southern Observatory (ESO).
The telescope is equipped
with the world's largest infrared camera, VIRCAM \citep{dalton06}
composed of 67 million pixels offering a field-of-view of 1.65 degrees in diameter,
0.6 square degrees of which is sampled by 'pawprints' of 16 non-contiguous
detectors with 0.34 arcsec pixels. Six suitably offsets and jittered pawprints are
combined into a filled 1.5 square degree 'tile', in which each piece of sky has
been sampled by at least 2 pixels (ignoring the jitters), together with two
additional 5.5$\times$88.5 arcmin strips at opposite sides of the tile which
are just covered once. The VISTA astrometric requirements, to an airmass of 2, 
are twofold: differential astrometric accuracies of 0.1 and
0.03 arcsec rms within the field covered by the full mosaic and each 
individual detector, respectively \citep{emerson04b}.

Seventy-five percent of the time available on VISTA is dedicated to European
Southern Observatory (ESO) public 
surveys\footnote{www.eso.org/sci/observing/PublicSurveys/sciencePublicSurveys.html},
one of them being the VHS \citep{mcmahon12}
planned to image the entire Southern sky in $J$ and $K_{s}$ over several years. 
Five thousands square degrees linked to the Dark Energy Survey 
(DES)\footnote{http://www.darkenergysurvey.org/} will 
be observed in $JHK_{s}$ whereas another 5000 square degrees around the 
Galactic caps will be imaged in $YJHK_{s}$. The averaged 5$\sigma$ depths 
measured for VHS are $J$\,=\,19.5$\pm$0.3 mag and $K_{s}$\,=\,18.5$\pm$0.3 mag.
One of the main scientific drivers of the VHS includes the search for and characterisation 
of the nearest and coolest low-mass stars and brown dwarfs. The current VHS 
data release is DR2\footnote{http://www.eso.org/sci/observing/phase3/data\_releases/vhs\_dr2.pdf},
made public on 17 October 2011 via the VISTA Science 
Archive \citep{hambly08,cross12}, but we also accessed data through flat-files 
made available to the VHS consortium.

%
%
\begin{table*}
{\tiny{
 \centering
 \caption[]{Wide common proper motion companions to planet-host
stars. We list their coordinates (J2000), $J$ and $K_{s}$ photometry
from VHS, $w1$ and $w2$ magnitudes from WISE, proper motions in
right ascension and declination (mas/yr) with a typical errors of 
15--20 mas/yr in each direction, the name of
the associated planet-host star with its spectral type in brackets,
the distance of the planet-host star, the separation in arcmin, 
proper motions of the primary (in mas/yr) with their errors from Hipparcos,
and the final conclusion (C?) on the validity of the system.
From top to bottom, panels list 
objects with ``Y'' which are truly wide common proper motion systems,
objects classified as ``Y:'' that are common proper motion candidates 
located at very wide separations, and rejected systems marked 
as ``N''.
}
 \begin{tabular}{@{\hspace{0mm}}c @{\hspace{1mm}}c @{\hspace{2mm}}c @{\hspace{2mm}}c @{\hspace{2mm}}c @{\hspace{2mm}}c @{\hspace{2mm}}c @{\hspace{2mm}}c @{\hspace{2mm}}c @{\hspace{2mm}}c @{\hspace{2mm}}c @{\hspace{2mm}}c @{\hspace{2mm}}c @{\hspace{2mm}}c @{\hspace{1mm}}c@{\hspace{0mm}}}
 \hline
R.A.  &  dec & $J$  & $K_{s}$  & $w1$  & $w2$  & $\mu_{\alpha}cos{\delta}$ & $\mu_{\delta}$ &  Name & dist & Sep & $\mu_{\alpha}cos{\delta}$ & $\mu_{\delta}$ & SpT & C? \cr
 \hline
hh:mm:ss.ss & $^{\circ}$:':''  & mag & mag & mag & mag & mas/yr & mas/yr  & & pc  & arcmin & mas/yr & mas/yr & \cr
 \hline
14:26:45.74 & $-$05:10:20.9 & 12.116 & 11.279 & 11.060 & 10.921 & $-$140 & $-$145 & HD\,126614  (K0V+MV) & 72.4  & 0.70 & $-$152.04$\pm$0.86 & $-$147.93$\pm$0.76 & M4.5$\pm$0.5 & Y \cr
14:28:42.32 & $-$46:28:41.7 & 15.533 & 15.582 & 15.174 & 13.853 & $-$35  & $-$196 & HIP\,70849  (K7Vk)   & 24.0  & 6.29 & $-$47.12$\pm$2.15 & $-$203.52$\pm$1.93 & T4.5$\pm$0.5 & Y \cr
22:31:08.09 & $-$49:26:59.0 & 12.415 & 11.595 & 11.208 & 11.030 & $-$143 & $-$191 & HD\,213240  (G0/G1V) & 40.6  & 1.59 & $-$136.01$\pm$0.48 & $-$194.44$\pm$0.33 & M5--M5.5 & Y \cr
 \hline
11:00:49.99 & $-$11:03:43.2 & 14.062 & 13.242 & 13.034 & 12.816 & $-$179 & $-$20  & BD\,10$-$3166 (K0V)  & 66.0  & 38.86 & $-$185.50$\pm$1.80 & $-$5.20$\pm$1.80 & M5--M5.5 & Y: \cr
12:37:18.37 & $-$39:53:14.8 & 12.350 & 11.542 & 11.046 & 10.899 & $-$185 & $-$9   & HD\,109749  (G3V)    & 56.3  & 55.48 & $-$157.85$\pm$0.93 & $-$5.43$\pm$0.72 & M3.5$\pm$0.5 & Y: \cr
15:58:41.31 & $-$44:27:54.2 & 14.568 & 13.749 & 13.591 & 13.324 & $-$176 & $-$79  & HD\,143361  (G6V)    & 65.66 & 33.76 & $-$158.32$\pm$1.37 & $-$118.21$\pm$1.01 & M6--M6.5 & Y: \cr
21:51:24.97 & $-$26:36:42.6 & 13.912 & 13.039 & 12.832 & 12.625 &    133 & $-$117 & HD\,207832 (G5V)     & 54.4  & 38.57 &    132.05$\pm$0.97 & $-$143.15$\pm$0.52 & M6.5$\pm$0.5 & Y: \cr
23:13:45.29 & $-$08:22:56.8 & 12.036 & 11.236 & 11.016 & 10.838 &    361 & $-$23  & 91\,Aqr     (K0III)  & 45.9  & 52.91 &    368.68$\pm$0.40 & $-$17.16$\pm$0.28 & M3.5$\pm$0.5 & Y:? \cr
 \hline
01:48:24.30 & $-$53:33:49.6 & 16.332 & 15.635 & 15.523 & 15.670 &    133 & $-$91  & HD\,10647   (F9V)    & 17.3  & 53.63 &    166.32$\pm$0.24 & $-$106.52$\pm$0.27 & --- & N \cr
11:57:01.82 & $-$12:45:39.0 & 16.939 & 15.645 & 14.700 & 14.296 & $-$165 &    7   & HD\,103774  (F5V)    & 55.0  & 39.20 & $-$125.50$\pm$0.49 & $-$19.95$\pm$0.34 & --- & N \cr
12:40:36.35 & $-$40:59:00.7 & 11.788 & 10.954 & 10.802 & 10.760 & $-$188 & $-$3   & HD\,109749  (G3V)    & 56.3  & 39.19 & $-$157.85$\pm$0.93 & $-$5.43$\pm$0.72 & K7 & N \cr
16:01:36.44 & $-$45:11:05.2 & 13.040 & 12.466 & 12.200 & 12.220 & $-$136 & $-$119 & HD\,143361  (G6V)    & 65.66 & 45.06 & $-$158.32$\pm$1.37 & $-$118.21$\pm$1.01 & Early & N \cr
17:36:35.14 & $-$42:35:39.8 & 13.378 & 12.724 & 12.639 & 12.527 & $-$194 & $-$147 & HD\,159868  (G5V)    & 58.7  & 42.34 & $-$231.62$\pm$1.04 & $-$167.00$\pm$0.81 & sdM0.5$\pm$0.5 & N \cr
17:41:02.53 & $-$42:34:54.8 & 12.769 & 12.097 & 11.912 & 11.732 & $-$214 & $-$175 & HD\,159868  (G5V)    & 58.7  & 40.69 & $-$231.62$\pm$1.04 & $-$167.00$\pm$0.81 & M3.0$\pm$0.5 & N \cr
22:53:23.80 & $-$47:54:01.8 & 12.631 & 11.881 & 11.660 & 11.467 &    215 & $-$89  & HD\,216435  (G0V)    & 32.62 & 41.95 &    217.04$\pm$0.37 & $-$81.76$\pm$0.29 & M4--M4.5 & N \cr
23:40:52.27 & $-$05:45:01.4 & 14.687 & 13.881 & 13.713 & 13.372 & $-$125 & $-$116 & HD\,222582  (G5 )    & 42.0  & 20.41 & $-$144.88$\pm$0.88 & $-$111.93$\pm$0.56 & --- & N \cr
 \hline
 \label{tab_compHIP:list_candidates}
 \end{tabular}
}}
\end{table*}
\subsection{Search criteria}
\label{compHIP:surveys_search}

We took advantage of the $>$10 year baseline between the VHS and 2MASS databases
\citep{mcmahon12,cutri03,skrutskie06} to search for wide companions to planet-host
stars imaged by the VHS\@. We searched for common proper motion companions to
planet-host stars over a separation of 60 arcmin. We identified a total of 
49 planet-host stars in the $\sim$6200 square degrees (assuming 1.2 square degree 
coverage for the 5175 files) covered by VHS up to 31 October 2012\@.
Among those 49 stars, 12 have incomplete coverage out to 60 arcmin at the time of 
writing, yielding a complete search for wide companions around 37 planet-host stars.
Our search probes a similar parameter space as the hunt for wide binaries 
in the rNLTT catalogue \citep{gould03a,salim03a} described by \citet*{chaname04},
except that we focus here on planet-host stars.
We applied the following criteria to our photometric and astrometric search:

\begin{itemize}
\item Sources brighter than $J_{\rm VHS}$\,=\,17 mag to ensure a reliable photometric detection in 2MASS
\item Difference in $J$-band magnitude between 2MASS and VHS photometry less than 0.5 mag to
remove unreliable photometry from 2MASS and VHS at the faint and bright ends, respectively
\item Sources with separations larger than 1 arcsec between 2MASS and VHS
\item Separations from the primary up to 60 arcmin
\item Differences of less than 40 mas/yr (typically 2--3$\sigma$ of the rms determined from stars
in the fields-of-view of our targets) between each component of the proper motion of the 
candidate(s) and the planet-host star, corresponding to a maximum total difference of 56 mas/yr
\end{itemize}

This query returned 16 wide co-moving proper motion candidates to 14 planet-host stars,
including one brown dwarf candidate with the typical colours of mid-T dwarfs
(Section \ref{compHIP:new_comp_BD}). 
We list the coordinates (J2000), near-infrared ($J,K_{s}$) photometry from VHS \citep{mcmahon12}, 
mid-infrared ($w1,w2$) photometry from the WISE all-sky release \citep{wright10}, proper 
motions in mas/yr for the primary and co-moving candidate from our VHS-WISE cross-match, 
physical separations, final spectral types (Section \ref{compHIP:new_comp}), and
conclusions on the true companionship of the candidates in Table \ref{tab_compHIP:list_candidates}.

\begin{figure}
  \centering
  \includegraphics[width=\linewidth, angle=0]{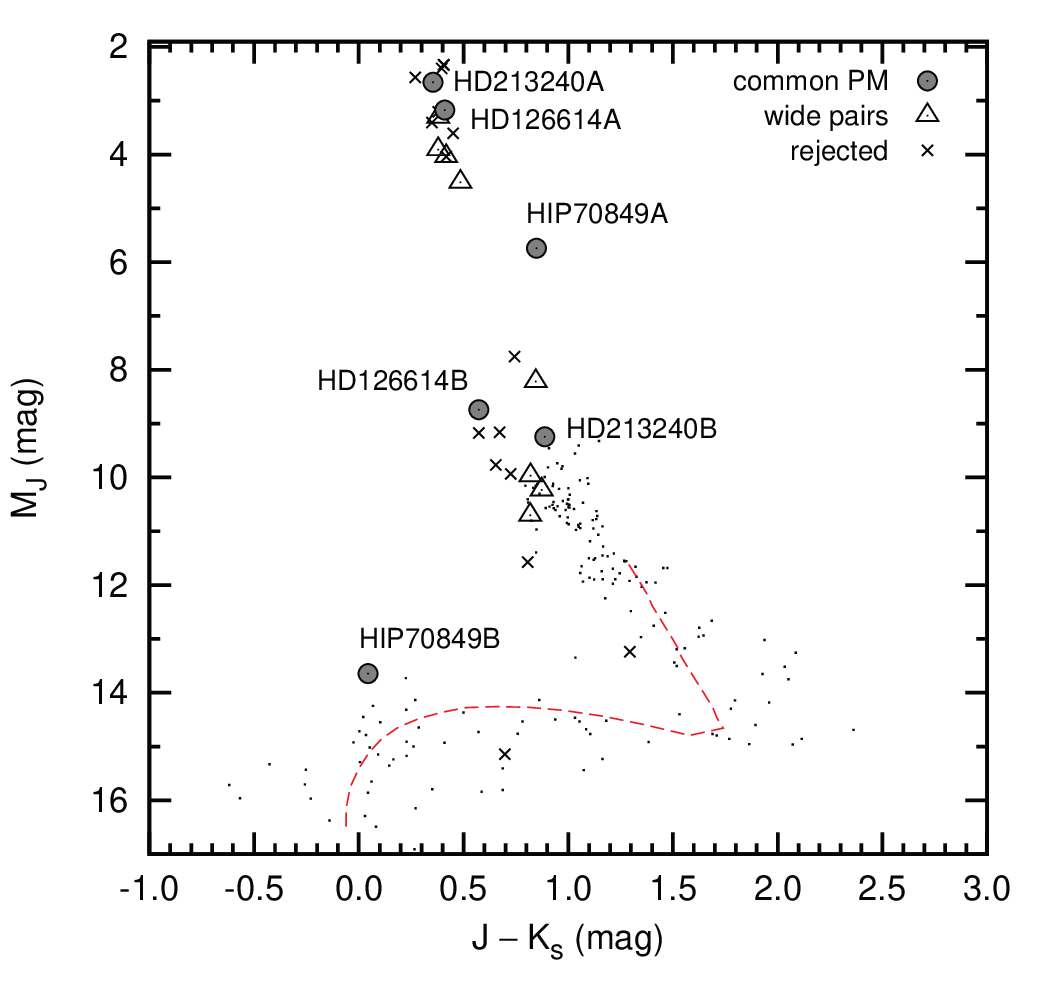}
   \caption{($J-K_{s}$,$M_{J}$) colour-absolute magnitude diagram for our wide co-moving
companion candidates to planet-host stars surveyed by VHS and 2MASS\@.
Filled circles indicate candidates confirmed spectroscopically as companions,
open triangles indicate very wide co-moving pairs, whereas crosses mark
rejected candidates.}
   \label{fig_compHIP:CMD_JKJ}
\end{figure}

We plot our candidates in the ($J-K_{s}$,M$_{J}$) colour-absolute magnitude diagram 
displayed in Fig.\ \ref{fig_compHIP:CMD_JKJ}, where we assumed the planet-host star 
distance to determine a possible absolute $J$-band magnitude.
We note that the companion to HD\,213240 was already published by \citet{mugrauer05c}
and that \citet{desidera07} already reported a wide companion to HD\,109749 but at a
closer separation.

We were able to discard three candidates photometrically before carrying out
the spectroscopic follow-up i.e.\ both components of the system don't seem
to be located at the same distance in the colour-absolute magnitude diagram
(e.g.\ Fig.\ \ref{fig_compHIP:CMD_JKJ}). We found these candidates around HD\,10647,
HD\,103774 (also discarded astrometrically after inspection of the images), 
and HD\,222582 (Table \ref{tab_compHIP:list_candidates}) but we do not 
discuss them further below. 

We estimate the typical dispersion observed in the vector point diagram for
the sources identified in 2MASS and VHS to be of the order of 15--20 mas.
In the case of three co-moving candidates to planet-host stars discussed
at length in Section \ref{compHIP:new_comp}, 
we derive dispersions along both axes of approximately 15--20 mas/yr
Moreover, we should take into account possible orbital motions, which we
estimate to be less than 9--12 mas over 12 years (typical baseline between 
2MASS and VHS). For our calculation, we considered again 
the case of the three confirmed companions (see Section \ref{compHIP:new_comp}). 
Their spectroscopic distances lie in the range 24--72 pc and their physical 
separations are larger than 3000 au. Moreover, we assumed longitudinal motions
(the worst case) with an amplitude of twice the separation over half of the 
period of the binary systems. Hence, we conclude that orbital motion is
negligible because its effect is below the uncertainty on our proper motion
determination. We note that \citet*{chaname04} found a general excess of 
proper motion differences for separations larger than 42 arcsec but for
closer systems.

%
%
\subsection{Known planet-host star binaries in VHS}
\label{compHIP:surveys_completeness}

We checked how many of the known planet-host stars within the VHS area 
considered in this work are known multiple systems in the literature to
gauge the completeness of our search. We identified a total of six wide 
or relatively wide systems which we briefly describe below.

\begin{enumerate}
\item HD\,109749 is a double G3IV star at 59 pc \citep{vanLeeuwen07} with a 
physical separation of approximately 750 au. The system harbours a 0.28 M$_{\rm Jup}$ 
planet with a period of 5.24 days \citep{nordstroem04,fischer06}.
\item HD\,126614 is a metal-rich (Fe/H\,=\,$+$0.56 dex) K0 star resolved in two 
components separated by 0.5 arcsec, corresponding to a projected physical 
separation of $\sim$36 au at a distance of 72.4$\pm$5.4 pc \citep{howard00a}.
A second M dwarf, NLTT,37349, has been reported as a wide companion by
\citet{gould04}.
\item HR\,6 (=\,HD\,142) is a double star at 20.6 pc \citep{vanLeeuwen07} whose components 
are separated by approximately 85 au \citep{fabricius02,gray06}. The star is orbited 
by three planets announced by \citet{tinney02} and \citet{wittenmyer12}.
\item HR\,8734 (=\,HD\,217107) is a G8IV double star at 19.72 pc \citep{montes01a,vanLeeuwen07}
made of the CDM\,J22583$-$0224A and CDM\,J22583$-$0224B components \citep{hog00,fabricius02}
with a separation of $\sim$10 au and orbited by two Jupiter-mass planets \citep{wright09a}.
\item HD\,28254 is a metal-rich (Fe/H\,=\,$+$0.36 dex) G1IV/V star at 56.2 pc \citep{vanLeeuwen07}.
It includes the components CCDM\,J04248$-$5037A; $J$\,=\,6.395 mag) and 
CCDM\,J04248$-$5037B ($J$\,=\,7.60 mag) separated by $\sim$240 au and a 
1.16 M$_{\rm Jup}$ planet with a long period \citep[P\,=\,1116 days;][]{naef10}
\item Gl\,86 is a metal-poor G9V star \citep{torres06} at 10.9 pc \citep{vanLeeuwen07}
with a white dwarf companion \citep[Gl\,86B;][]{els01,mugrauer05a,lagrange06} at a 
separation of $\sim$17.4 au. \citet{queloz00a} announced a 4 M$_{\rm Jup}$ planet 
orbiting Gl\,86 in $\sim$15.8 days.
\end{enumerate}

We do not recover any of these planet-host stars because each component is very bright
($J_{\rm 2MASS}$\,=\,4.7--8.8 mag), leading to differences in the $J$-band magnitudes 
between 2MASS and VHS larger than 0.5 mag, our upper limit (Section \ref{compHIP:surveys_search}).
This effect comes from the heavy saturation effects seen for these stars on the VHS
images. Furhermore, HR\,8734 has a very low proper motion in both directions whereas
the mean proper motions of HD\,109749 and HD\,28254 are slightly larger than 0.15 arcsec/yr.
Moreover, HD\,126614 is unresolved on the seeing-limited VHS images.

%
%
\section{Spectroscopic follow-up of potential companions}
\label{compHIP:spectro}
\subsection{NTT/EFOSC2 optical spectroscopy}
\label{compHIP:spectro_opt_EFOSC2}

We obtained optical spectroscopy of eight of the ten new systems with the
ESO Faint Object Spectrograph and Camera \citep[EFOSC2;][]{buzzoni84} installed 
on the 3.58-m New Technology Telescope (NTT) in La Silla (Chile) on 11 August 2013
in visitor mode (Table \ref{tab_compHIP:log_obs}). The observations of our candidates 
took  place between UT\,=\,0h and 5h. The night was non photometric with thin cirrus 
all night. The moon was illuminated at 48\% and was up until UT\,=\,5h. The seeing 
oscillated between 1 and 1.7 arcsec during that night.
We observed three other candidates with the same set-up on 11 March 2014, a clear night
with photometric conditions and seeing between 0.7 and 0.9 arcsec. We observed
our targets at the end of the night when the moon had set.

EFOSC2 is equipped with a Loral/Lesser, thinned, ultra-violet flooded 2048$\times$2048
chip sensitive to optical wavelengths. The 2$\times$2 binning mode offers a pixel scale of
0.24 arcsec and a field-of-view of 4.1$\times$4.1 arcmin.
We used grism \#5 with a slit of 1 arcsec for spectroscopic observations, covering the
520--935 nm wavelength range with a resolution of 1.54 nm. We obtained single or multiple
on-source integrations between 300\,sec and 1800\,sec at parallactic angle, depending 
on the brightness of the object (Table \ref{tab_compHIP:log_obs}). We observed 
internal flat fields straight after the spectra of the targets to correct for fringing 
at red wavelengths. We also obtained bias and Helium-Argon arc frames during the 
afternoon preceding our observations as well as one spectro-photometric
standard star per run observed in between our targets to calibrate them in flux
\cite[LTT7379 (G0); HR\,5501 (B9.5][]{vanAltena95,gontcharov06,vanLeeuwen07,pancino12} .

We carried out the data reduction of all optical spectra under the IRAF environment
\citep{tody86,tody93}\footnote{IRAF is distributed by the National Optical Astronomy
Observatories, which are operated by the Association of Universities for Research in
Astronomy, Inc., under cooperative agreement with the National Science Foundation}.
We subtracted the raw median-combined bias frame from the raw spectrum and divided
by a normalised internal flat field obtained at the same position of our
targets to minimize the fringing in the red part of the spectrum.
We extracted a one-dimensional (1D) spectrum, choosing an optimal sky background and aperture.
In the case of multiple exposures, we combined the spectra before conducting the extraction.
We calibrated our spectrum in wavelength with Helium-Argon arc lamps before calibrating it
in flux, using the spectro-photometric standard star. The final optical spectra of
our targets, normalised at 750 nm, are displayed in Figure \ref{fig_compHIP:OPTspectrum}.

%
%
\begin{table}
 \centering
 \caption[]{Log of the spectroscopic observations. We give two integration times 
 for the SofI observations because we observed the object with the blue and red
 gratings to cover the full 0.9 to 2.5 micron wavelngth range.}
 \begin{tabular}{@{\hspace{0mm}}c @{\hspace{2mm}}c c c c@{\hspace{0mm}}}
 \hline
R.A.  &  dec & Tel.\ Inst.\ & Date & ExpT \cr
 \hline
hh:mm:ss.ss & $^{\circ}$:':''  &  & dd/mm/yy & sec \cr
 \hline
01:48:24.30 & $-$53:33:49.6 & no spectrum &  ---      & --- \cr
11:00:49.99 & $-$11:03:43.2 & GTC OSIRIS & 04/11/13 & 240 \cr
11:57:01.82 & $-$12:45:39.0 & no spectrum &  ---      & --- \cr
12:37:18.37 & $-$39:53:14.8 & NTT EFOSC2  &  11/03/14 & 4$\times$300 \cr
12:40:36.35 & $-$40:59:00.7 & NTT EFOSC2 &  11/08/13 & 300 \cr
14:26:45.74 & $-$05:10:20.9 & NTT EFOSC2  &  11/03/14 & 3$\times$600 \cr
14:28:42.31 & $-$46:28:41.7 & NTT SofI & 29/12/12 & 960$+$1200 \cr
15:58:41.31 & $-$44:27:54.2 & NTT EFOSC2 &  11/08/13 & 1800 \cr
16:01:36.44 & $-$45:11:05.2 & NTT EFOSC2  &  11/03/14 & 3$\times$600 \cr
17:36:35.14 & $-$42:35:39.8 & NTT EFOSC2 &  11/08/13 & 1200 \cr
17:41:02.53 & $-$42:34:54.8 & NTT EFOSC2 &  11/08/13 & 600 \cr
21:51:24.97 & $-$26:36:42.6 & NTT EFOSC2 &  11/08/13 & 1800 \cr
22:31:08.09 & $-$49:26:59.0 & NTT EFOSC2 &  11/08/13 & 600 \cr
22:53:23.80 & $-$47:54:01.8 & NTT EFOSC2 &  11/08/13 & 600 \cr
23:13:45.29 & $-$08:22:56.8 & NTT EFOSC2 &  11/08/13 & 600 \cr
23:40:52.27 & $-$05:45:01.4 & no spectrum &  ---      & --- \cr
 \hline
 \label{tab_compHIP:log_obs}
 \end{tabular}
\end{table}
\subsection{GTC/OSIRIS optical spectroscopy}
\label{compHIP:spectro_opt_GTC}

We obtained low-resolution (R\,=\,500) optical spectroscopy with the OSIRIS
\citep[Optical System for Imaging and low-intermediate Resolution Integrated Spectroscopy;][]{cepa00}
mounted on the 10.4-m GTC telescope in La Palma.  We observed the potential 
companion to BD\,10--3166 on 4 November 2013 in service mode 
(Table \ref{tab_compHIP:log_obs}) as part of program GTC65-13B (PI Lodieu).

OSIRIS is equipped with two 2048$\times$4096 Marconi CCD42-82 detectors 
offering a field-of-view approximately 7$\times$7 arcmin with an unbinned 
pixel scale of 0.125 arcsec.
The night was dark with spectroscopic conditions and a seeing of 1.7 arcsec.
We used a slit of 1.0 arcsec, the R500R grating and a single on-source
integration of 240\,sec. 
Bias frames, dome flat fields, and Neon, Xenon, and HgAr arc lamps were observed 
by the observatory staff during the afternoon preceding the observations.

\begin{figure}
  \centering
  \includegraphics[width=\linewidth, angle=0]{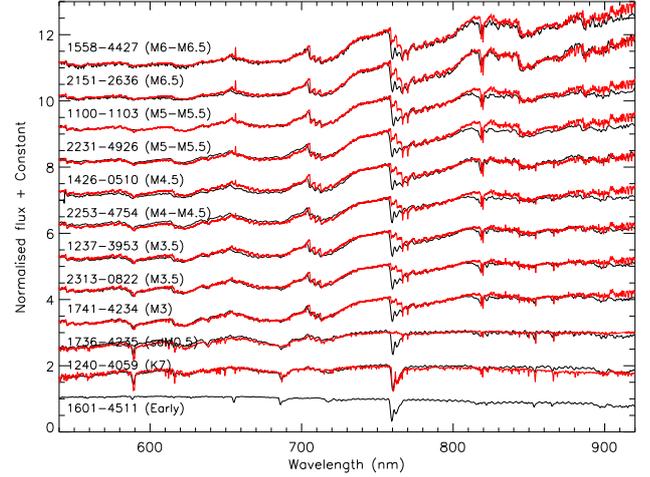}
   \caption{Low-resolution optical spectra for wide companion
   candidates obtained with NTT/EFOSC2 and GTC/OSIRIS 
   (Table \ref{tab_compHIP:log_obs}). The names of the candidate companions 
   and final spectral types are written on top of each spectrum. Spectra
   are ordered by increasing spectral types.}
   \label{fig_compHIP:OPTspectrum}
\end{figure}

We reduced the OSIRIS optical spectrum in a very similar way as the EFOSC2
spectra. We substracted the raw spectrum by a median-combined bias and
divided by a normalised median-combined dome flat field taken during the
afternoon. We extracted optimally a 1D spectrum from the 2D image and
calibrated that spectrum in wavelength with the lines from the combined arc lamp.
We calibrated our spectrum in flux with the GD\,248 spectro-photometric standard 
\citep{farihi05,landolt07} corrected for the second-order contamination.
We corrected the second-order by joining the response of the detector
from the observations of the standard star with the R500R grating and
the same grating in combination with the Sloan $z$ broad-band filter.
The final GTC spectrum of 2MASS\,11004999$-$1103432, normalised at 750nm, 
is displayed in Figure \ref{fig_compHIP:OPTspectrum}.

%
%
\subsection{NTT/SofI near-infrared spectroscopy}
\label{compHIP:spectro_NIR_SofI}

We performed near-infrared spectroscopy of the wide companion to HIP\,70849
on 29 December 2012 with the Son of ISAAC (SofI) instrument mounted on the 
3.5-m New Technology Telescope \citep[NTT;][]{moorwood98} in La Silla, Chile
(Table \ref{tab_compHIP:log_obs}). 
SofI is equipped with a Hawaii HgCdTe 1024$\times$1024 array with squared 
18.5 micron pixels and offers imaging and spectroscopic capabilities. The pixel 
scale is 0.292 arcsec in the Large Field configuration, offering a 4.9$\times$4.9 arcmin
field-of-view. We observed our target with the blue $ZJ$ and red $HK$ gratings,
covering 0.9--1.64 and 1.53--2.52 microns, respectively, at a resolution of
$\sim$600 with a slit of 1 arcsec. We set one and two ABBA patterns with individual 
on-source integrations to 60 sec repeated four and five times in the $ZJ$ and $HK$ gratings, 
respectively, to correct for the sky. The total integration times are 20 min and 40 min for 
the blue and red gratings, respectively. We observed a telluric standard immediately 
after our target at a similar airmass to correct for telluric absorption,
HIP\,68930 \citep[A0V;][]{houk78,vanLeeuwen07}.
Dome flats and halogen arc lamps were taken during the afternoon preceding the
beginning of the night with the same set-up as our observations. 
 
\begin{figure}
  \centering
  \includegraphics[width=\linewidth, angle=0]{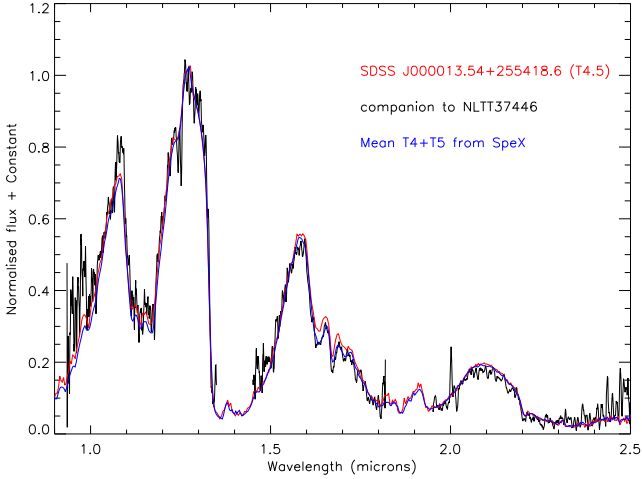}
   \caption{NTT/SofI low-resolution near-infrared spectra of the T4.5 dwarf 
co-moving with HIP\,70849 (black line). Overplotted are IRTF/SpeX spectra of 
SDSS J000013.54$+$255418.6 \citep[T4.5;][]{knapp04,jameson08a,burgasser06a}
and the average spectrum of a T4 and T5 dwarf.
}
   \label{fig_compHIP:NIRspectrum}
\end{figure}

Each individual frame was flat-fielded and sky-subtracted before a one-dimensional
spectrum was extracted. We combined the individual one-dimensional spectra
to increase the signal-to-noise. Then, we divided this combined spectrum of
the target by the extracted spectrum of the standard star and later 
multiplied by an appropriate spectral template of spectral type A0 (containing
the main hydrogen lines in absorption) smoothed to our resolution.
The final one-dimensional spectrum of HIP\,70849B, normalised at 1.25 microns, is shown 
in Figure \ref{fig_compHIP:NIRspectrum} along with known T dwarf spectral standards
taken from the IRTF/SpeX library\footnote{http://pono.ucsd.edu/$\sim$adam/browndwarfs/spexprism/}.

%
%
%

%
%
\section{Optical spectral classification}
\label{compHIP:optical_SpT}

To classify the components of our potential wide multiple systems, we opted
for the direct comparison of our optical spectra with known M dwarf templates 
created from the Sloan Digital Sky Survey \citep[SDSS;][]{york00} spectroscopic 
database by \citet{bochanski07a}. This database contains a repository of 
good-quality M0--M9 spectra spanning the 380--940 nm wavelength range at a resolution 
of 2000\@. All spectra are wavelength- and flux-calibrated and corrected for 
telluric absorption. M dwarfs are classified based on the Hammer classification 
scheme \citep{covey07} which uses the spectral energy distribution of stars over 
the 0.3--2.5 micron range using photometry from SDSS and 2MASS\@.
We added sub-type to this list of templates to provide a complete list of
spectral templates with optical spectral types accurate to one sub-type.

In addition to this list of solar-metallicity M dwarf templates, we dowloaded
the spectra of the brightest object of each sub-class of low-metallicty M dwarfs,
classified either as subdwarfs, extreme subdwarfs, or ultrasubdwarfs.
The low-metallicity classification follow the scheme by \citet{lepine07c},
which is an updated version of the original scheme proposed by \citet{gizis97a}.
For more details, we refer the reader to Espinoza Contreras et al.\ (2014).

We compared each of our optical spectra to this grid of spectral templates
and assign a spectral type to the nearest half sub-class, which we consider
as our (absolute) uncertainty. All our targets are either solar-type M dwarfs
or M subdwarfs, except 2MASS\,J12403635$-$4059007 which we classify as a K7
dwarf (Table \ref{tab_compHIP:list_candidates}).

%
%
\section{Absolute magnitude vs spectral type relation}
\label{compHIP:AbsMag_SpT}

Throughout this section and the rest of the paper, we will make use of the absolute magnitude 
vs spectral type relations given in Table \ref{tab_compHIP:dM_absMag} to derive 
spectroscopic distances for the co-moving companion candidates to planet-host stars. 
We calculated the absolute $J$-, $H$-, and $K_{s}$-band magnitudes using the 8 pc 
census of M dwarfs listed in Table 4 of \citet{kirkpatrick12}. We compiled all M dwarfs 
with spectral types between M0 and M7 with parallaxes and accurate spectral types. 
We computed the average values (Table \ref{tab_compHIP:dM_absMag}), after excluding 
some outliers and members of multiple systems with unresolved 2MASS photometry. 
We also list in brackets the dispersion in absolute magnitudes taken from the full
sample of objects considered for each sub-type.

%
%
\begin{table}
 \centering
 \caption[]{Absolute $J$, $H$, and $K_{s}$ magnitudes as a function of
spectral types for M dwarfs with spectral types between M0 and M7 drawn
from the census of 8 pc compiled by \citet{kirkpatrick12}. The last
column indicates the number of stars used to derive the average absolute
magnitudes}
 \begin{tabular}{@{\hspace{0mm}}c @{\hspace{2mm}}c @{\hspace{2mm}}c @{\hspace{2mm}}c c@{\hspace{0mm}}}
 \hline
SpType &  M$_{J}$ & M$_{H}$ & M$_{K_{s}}$ & Nb \cr
 \hline
       &  mag (err)    &   mag   &   mag       &     \cr
 \hline
M0.0   &  6.273 (0.322)  &   5.577 (0.440) &   5.408 (0.337)  &  3  \cr
M0.5   &  6.761 (0.258)  &   6.031 (0.230) &   5.888 (0.200)  &  1  \cr
M1.0   &  6.393 (0.089)  &   5.736 (0.144) &   5.491 (0.125)  &  2  \cr
M1.5   &  6.943 (0.540)  &   6.350 (0.572) &   6.063 (0.537)  &  8  \cr
M2.0   &  7.134 (0.584)  &   6.554 (0.608) &   6.253 (0.597)  &  4  \cr
M2.5   &  7.255 (0.288)  &   6.667 (0.297) &   6.409 (0.294)  &  5  \cr
M3.0   &  7.758 (0.650)  &   7.206 (0.664) &   6.936 (0.645)  &  9  \cr
M3.5   &  8.025 (0.458)  &   7.460 (0.454) &   7.183 (0.464)  & 14  \cr
M4.0   &  8.558 (0.518)  &   8.027 (0.567) &   7.721 (0.563)  & 10  \cr
M4.5   &  9.038 (0.420)  &   8.495 (0.440) &   8.183 (0.438)  &  7  \cr
M5.0   &  9.617 (0.121)  &   9.029 (0.164) &   8.685 (0.136)  &  3  \cr
M5.5   &  9.671 (0.392)  &   9.101 (0.397) &   8.745 (0.394)  &  7  \cr
M6.0   & 10.439 (0.025)  &   9.883 (0.069) &   9.535 (0.120)  &  2  \cr
M6.5   & 10.505 (0.067)  &   9.899 (0.079) &   9.552 (0.089)  &  2  \cr
M7.0   & 10.776 (0.026)  &  10.126 (0.022) &   9.739 (0.027)  &  1  \cr
 \hline
 \label{tab_compHIP:dM_absMag}
 \end{tabular}
\end{table}

%
%
\section{New wide companions to planet-host stars}
\label{compHIP:new_comp}

In this section, we only describe the properties of the three planet-host 
stars with most probable wide companions and derive their main properties
(Table \ref{tab_compHIP:list_candidates})
thanks to our spectroscopic follow-up (Section \ref{compHIP:spectro}).
The other companions, located at very wide separations or rejected as co-moving
companions to planet-host stars are discussed in the appendix 
\ref{compHIP:new_comp_appendix}.

\subsection{HD\,126614}
\label{compHIP:HD126614}

\citet{howard00a} announced a 0.38 M$_{\rm Jup}$ planet with a period of 1244 days
orbiting the metal-rich (Fe/H\,=\,$+$0.56 dex) K0 star, HD\,126614\@.
These authors also resolved HD\,126614 in two components separated by 0.5 arcsec
\citep{howard00a}, corresponding to a projected physical separation of $\sim$36 au 
at a distance of 72.4$\pm$5.4 pc \citep{vanLeeuwen07}. However, if we apply
a spectral type vs absolute magnitude relation for K0 main-sequence stars using
Gl\,68 as a K0 template at 7.53 pc \citep[$J$\,=\,3.83 mag;][]{gray06,kirkpatrick12},
we derive a spectroscopic distance of $\sim$40 pc for HD\,126614\@. This discrepancy
might arise from two facts. On the one hand, HD\,126614 is not a main-sequence K0 star 
but rather a metal-rich K0 sub-giant because of its temperature (5541--5645\,K) and 
mass \citep[1.03--1.25 M$_{\odot}$;][]{howard00a,santos13}. On the other hand, 
HD\,126614 is a metal-rich star, with [Fe/H]\,=\,0.46--0.60 \citep{howard00a,santos13}.
Hence, the above relationship for solar-type K0 main-sequence star is not applicable
so we favour the Hipparcos distance.

We identified a relatively close companion at 41.8 arcsec from the planet-host
system, previously reported by \citet{gould04} as being NLTT\,37349\@. 
We confirm the common proper motion of the pair after examination of the 
optical and infrared images of the system taken by the Digital Sky Survey, 
2MASS, and VHS over a period of more than 50 years. The common motion of
the two sources over such long period adds strong credibility to their association.

We obtained an optical spectrum of this close companion and classify it as 
M4.5$\pm$0.5 dwarf. If a true companion, this M4.5 dwarf represents a
very interesting M dwarf with super-solar metallicity that is amenable for 
further investigation.
We derive a spectroscopic distance of 41.3 pc with a possible range of 29--65 pc,
assuming that the absolute magnitude vs spectral type relation is valid for 
(single) solar-metallicity M dwarfs. We note that, according to the Padova tracks 
\citep{bertelli08},
a 0.16 M$_{\odot}$ dwarf is less luminous ($\sim$0.07 dex), cooler ($\sim$200\,K),
and fainter by approximatly 0.1 mag in the near-infrared than a solar-metallicity
object with the same mass. Moreover, if this companion is an equal-mass binary,
it would lie at a distance greater by 40\% (41--91 pc).
This value is in agreement with the spectroscopic distance of the primary
and the range of possible distances is within 1.5$\sigma$ of the Hipparcos distance.
Hence, we argue that these two objects form a physical pair with a projected physical
separation of approximately 3000 au (assuming a distance of 72.4 pc).

\subsection{HD\,213240}
\label{compHIP:HD213240}

\citet{santos01b} reported an eccentric (0.45$\pm$0.05) planet with
a mass of 4.5 M$_{\rm Jup}$ and a period of $\sim$950 days, orbiting 
HD\,213240, an old (5.11 Gyr) G0/G1V star \citep{houk78} at 
40.6$\pm$1.0 pc \citet{vanLeeuwen07} and with a metallicity of 
[Fe/H]\,=\,0.16 \citep{nordstroem04}.

We unveiled a potential wide companion to HD\,213240, located at
1.6 arcmin only, the closest companion found among our candidates.
We classify this companion as a M5--M5.5 dwarf, translating into a
spectroscopic distance of 29--42 pc.

We conclude that these two objects form a co-moving pair at a projected physical
separation of $\sim$3900 au, adopting the distance of the primary.

%
%
\begin{figure*}
  \centering
  \includegraphics[width=0.46\linewidth, angle=0]{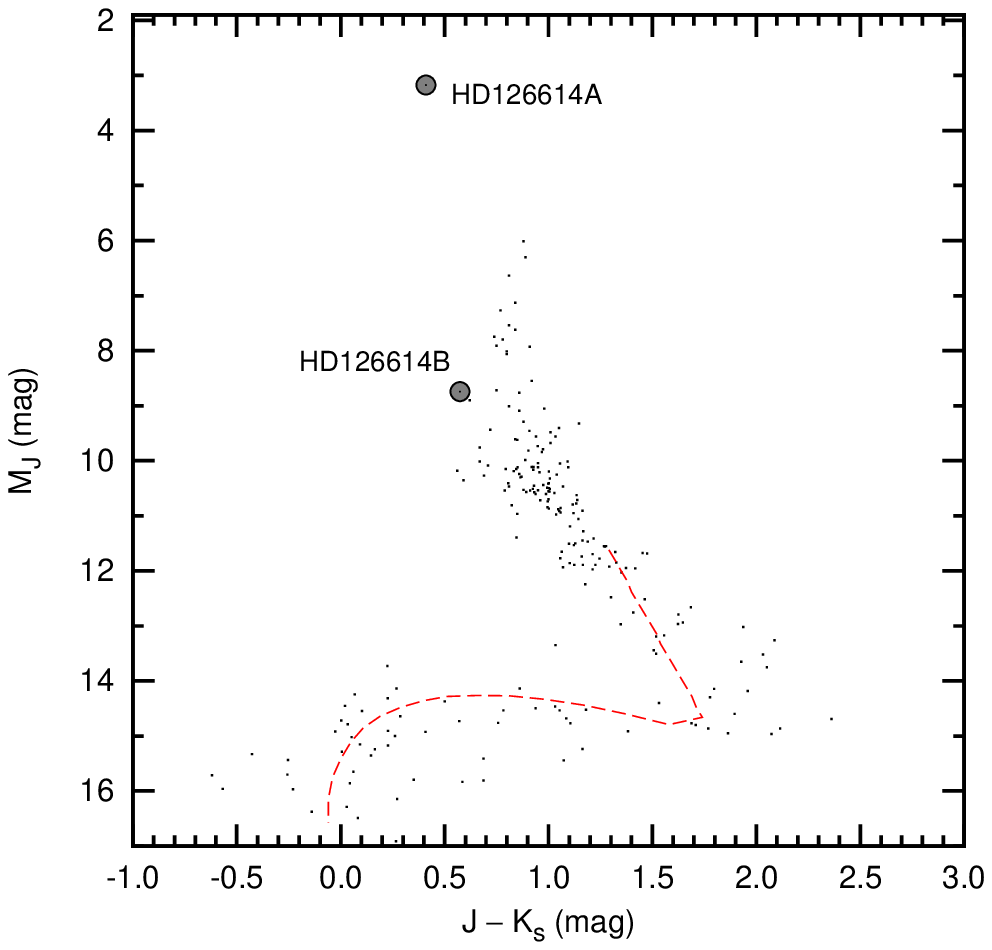}
  \includegraphics[width=0.46\linewidth, angle=0]{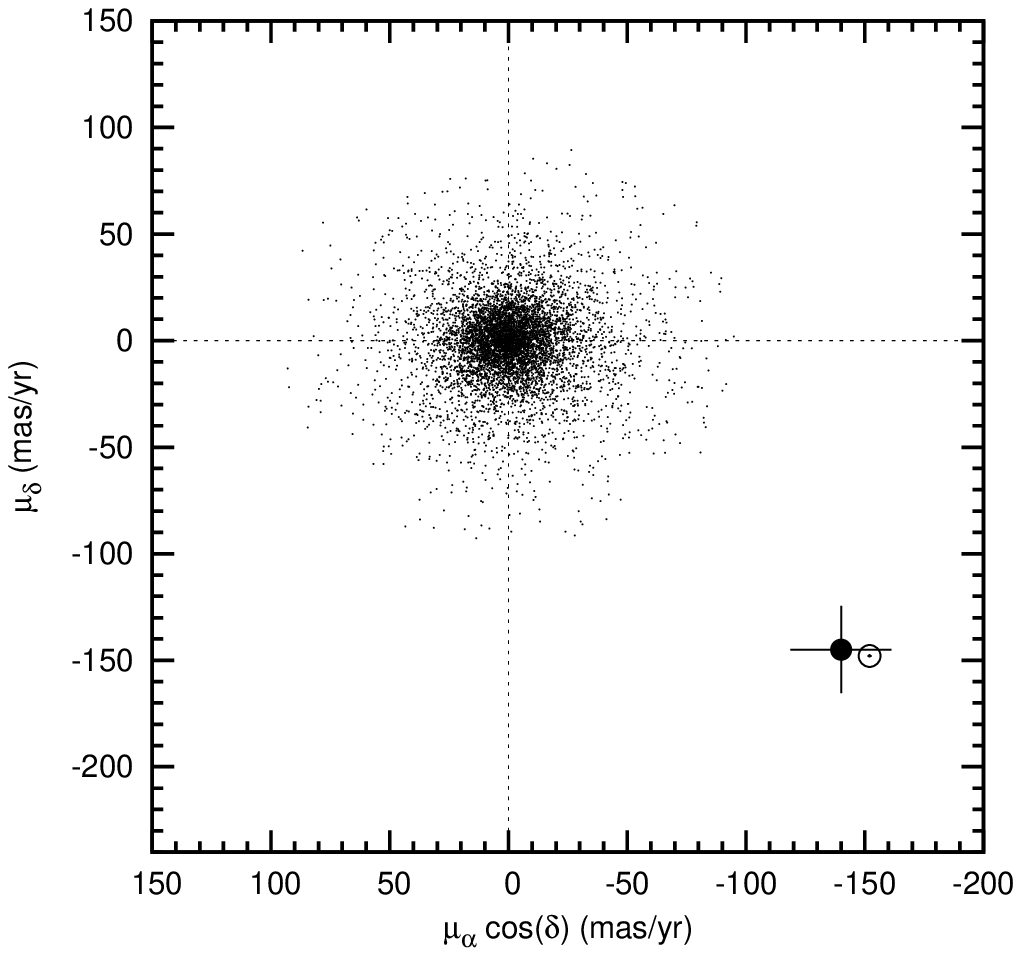}
  \includegraphics[width=0.46\linewidth, angle=0]{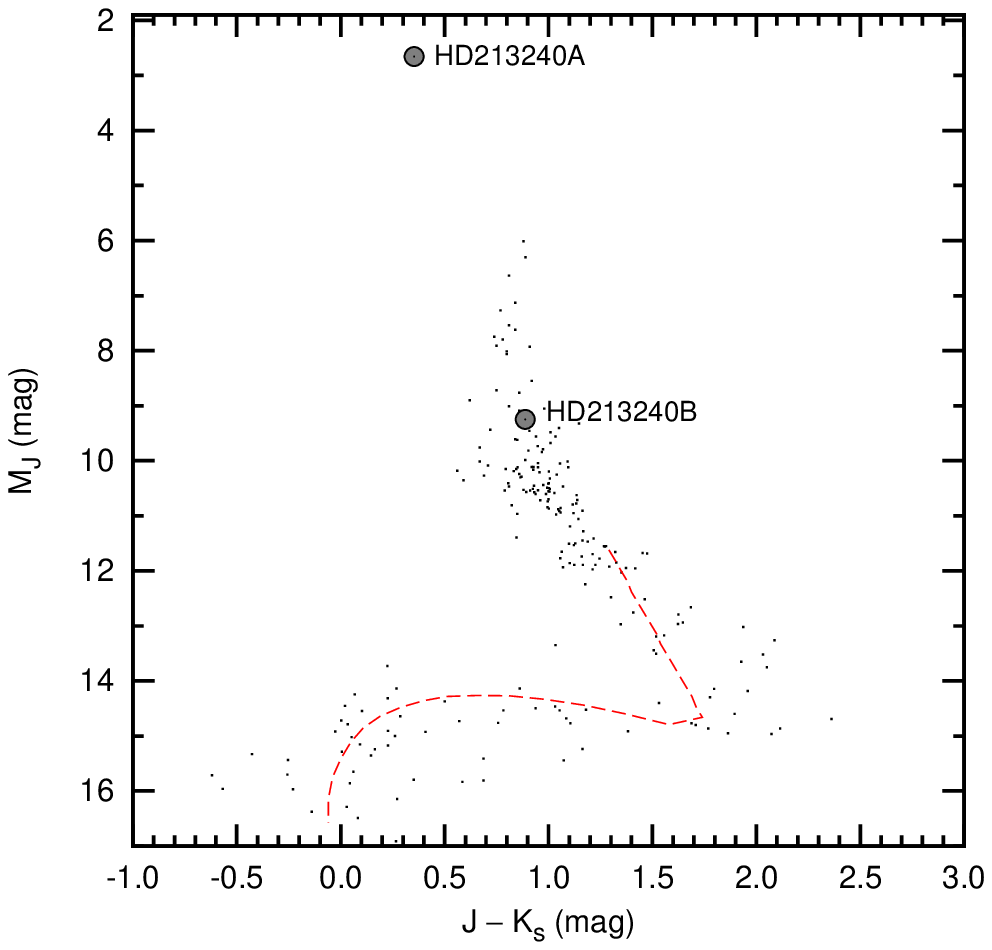}
  \includegraphics[width=0.46\linewidth, angle=0]{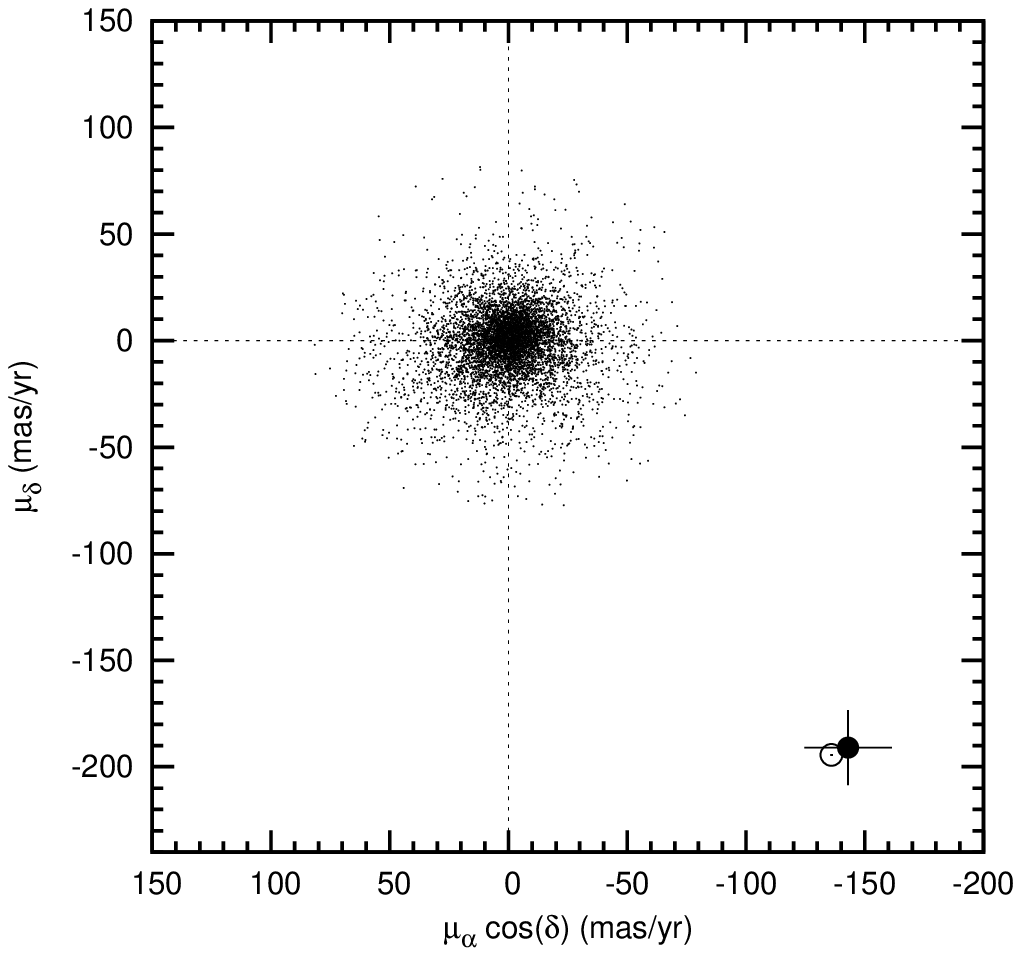}
  \includegraphics[width=0.46\linewidth, angle=0]{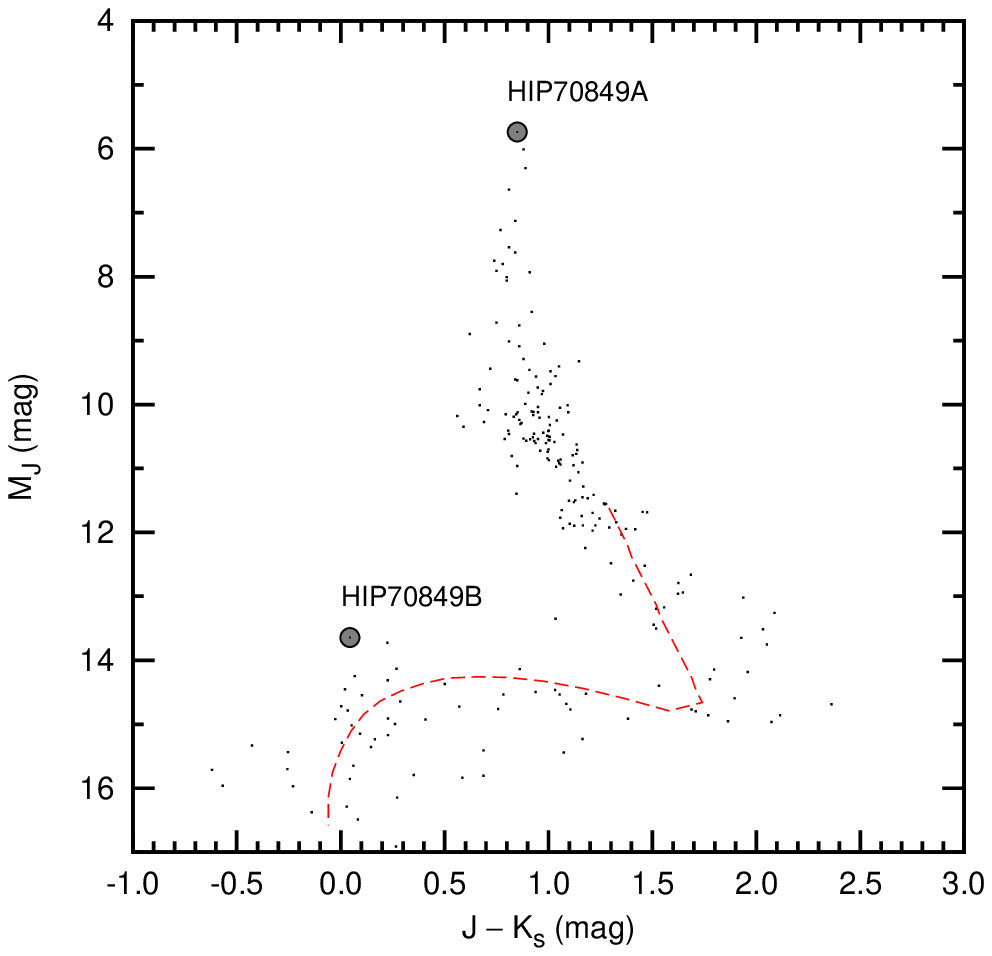}
  \includegraphics[width=0.46\linewidth, angle=0]{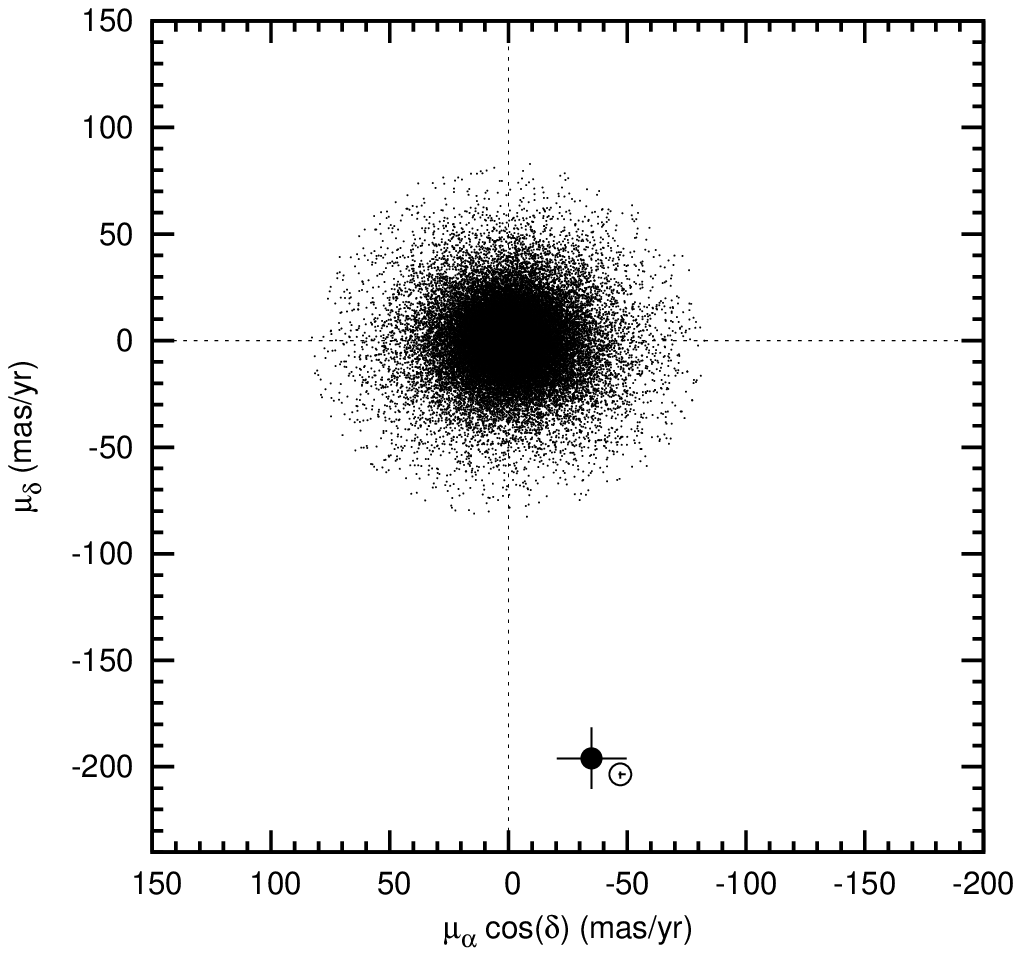}
   \caption{
{\it{Left:}} Proper motion diagram for all point sources within
60 arcmin from the primary stars with confirmed wide common proper
motion companions: HD\,126614 (top), HD\,213240 (middle), HIP\,70849 (bottom).
{\it{Right:}} Colour-magnitude diagrams showing the position of the three
planet-host stars and their companions. Overplotted are known M, L and T dwarfs
with parallaxes from \citet{leggett00a}, \citet{vrba04}, and \citet{kirkpatrick12}.
}
   \label{fig_compHIP:BDcomp}
\end{figure*}
\subsection{A new brown dwarf companion to HIP\,70849}
\label{compHIP:new_comp_BD}

\citet{segransan11} reported a 9$\pm$6 M$_{\rm Jup}$ planet orbiting 
HIP\,70849 in a very eccentric orbit (0.60$\pm$0.13) associated to
a long period of $\sim$10000 days. The planet-host star is a low-mass 
star (0.63$\pm$0.03 M$_{\odot}$) of spectral type K7V located 
24.0$\pm$1.2 pc \citep{vanLeeuwen07} from the Sun, an age of 1--5 Gyr, 
and a metallicity possibly close to solar.

We detected a potential wide proper motion companion at 
6.3 arcmin from HIP\,70849\@. We unambiguously confirm the
association of the potential brown dwarf companion and the
planet-host in the vector point diagram and in the 
($J-K$,M$_{J}$) colour-absolute magnitude diagram displayed in
Fig.\ \ref{fig_compHIP:BDcomp}. We obtained near-infrared
spectroscopy with NTT/SofI (Section \ref{compHIP:spectro_NIR_SofI}),
classifying the companion as a T4.5 dwarf 
(Fig.\ \ref{fig_compHIP:NIRspectrum}), hence, confirming
its substellar status. It is the second brown dwarf directly imaged
around a planet-host star after HD\,3651 \citep{mugrauer06b} and
possibly the first brown dwarf {\it{binary}} at wide separation from a
star orbited by a 9 M$_{\rm Jup}$ planet.

To check whether the binary is physically bound or a chance 
alignment, we have estimated the probability that the companion 
is a field T dwarf contaminant in our search. First, we have 
estimated that the expected number of T dwarfs in the full area 
of our search ($\sim$\,150 deg$^2$ around the full sample of 49 stars) 
is $\sim$\,1.3\@. We have adopted a density of 5$\times$10$^{-3}$\,objects/pc$^{-3}$ 
\citep{burgasser03f} and an average absolute magnitude of M$_{J}$\,$\sim$\,15 mag
\citep{vrba04} for such objects, and considered that our search is 
limited to $J$\,=\,17 mag. Secondly, we have estimated the probability 
that one of this object has a proper motion within 40 mas/yr of
the primary's proper motion in each direction. We have determined 
that about 20\% of the high proper motion objects in our full
catalogue have such similar proper motion as HD\,70849\@.
Hence, we estimate that the final probability that a field T dwarf 
lies at separations less than 60 arcmin of one of our primaries is 
of 0.27\@. If we take into account that the T4.5 was found at a 
separation of 6.29 arcmin from the primary, the probability of 
contamination decrease by a factor of 100, yielding a probability of 0.027\@. 
These estimates combined with the similar spectroscopic distances derived 
for both components support that the T4.5 is a true physical companion 
of the HD\,70849 planet-host star. 

%
%
\section{Stability of wide companions}
\label{compHIP:stability}

Three of our planet-host star companions (HIP\,70849\,B, HD\,213240\,B, HIP\,126614\,B) 
are located at angular separations between 0.7 and 6.3 arcmin, corresponding to physical 
separations in the 3000--9100 au range. The mass of their primaries lie between 0.63 and 
1.22 M$_{\odot}$. To study the dynamical stability of these systems, we can 
investigate their separations versus total masses \citep[see Fig.\ 16 in][]{close07a}.
From this kind of diagrams, we can see that these three systems are located in the 
region that is dynamically stable for the typical stellar densities of the Solar vicinity.

However, the other planet-host star companions are located at physical separations of 
100,000--150,000 au (0.50-0.75 pc). Given that the mass of their primaries are about 
1 M$_{\odot}$, these extremely wide binary systems are not dynamically stable at the 
typical density of field stars and the age of the Solar vicinity but do exist
\citep{scholz08c,shaya11}. Although binaries with separations larger than 10,000 au 
and up to 200,000 au have been reported in the literature
\citep[e.g.][]{makarov08,quinn09,poveda09,raghavan10,deacon12b},
most of these systems are either young or expected to be disrupted at older ages 
or are halo binaries. 
Another possibility is that these systems could be disrupted binaries that drift
apart quite slowly because disruption does not necessarily denote a violent phenomenon, 
suggesting that these pairs would survive as common proper motion associations for much 
longer than their disruption time-scales \citep{shaya11}.
Following the analytical solutions given in \citet{weinberg87} and \citet{binney87}
we estimate the timescale for disruption of these systems in an environment with typical 
densities of the solar vicinity ($\sim$ 0.07 object\,pc$^{-3}$) of a several hundred 
million years. This is much lower than the expected age of these systems, considering 
that we have not found any evidence of youth in the spectra of the secondaries. 
These estimations suggest that these systems are not physically bound. Given the large 
uncertainties in the determination of their distances, they are either objects with 
similar proper motions that lie along the same line of sight by chance, but lie at 
different distances, or they are systems located at the same distance and share similar 
projected velocities. This may be the result of a common origin or a similar dynamical 
evolution, in  the same way as stars which are members to moving groups.

We have estimated the probability of these five wide companions to be truly
associated with the planet-host stars. We counted the numbers of high proper motion
stars in the full 2MASS-VHS catalogue with similar motions as the planet-host star.
We found 19553 objects that fulfill those criteria, implying that we would expect
9.9$\pm$3.1 contaminants (assuming Poisson errors) around the 49 planet-host stars 
located in 6200 square degrees surveyed by the VHS (Table \ref{tab_compHIP:list_candidates})..
However, this number represents an upper limit because we have added a distance
criterion to our sample, based on our spectroscopic follow-up.
We rejected eight of our candidates, in agreement with the minimum of seven 
contaminants. Additionally, we have five very wide companions sharing the same 
proper motion and distance as the planet-host stars. We conclude that among the 
13 doubtful candidate companions in Table \ref{tab_compHIP:list_candidates} a few could
be part of co-moving groups. We have three very wide companions
with spectral types later than M5, whose number of field objects is typically lower
than earlier spectral types, suggesting that those are most likely to be associated 
to th  planet-host star (if any of these are). This result is not so surprising
and in line with the conclusions of \citet*{chaname04} who argue that binaries 
with separations larger than 8.3 arcmin belonging to the thin disk can not be 
identified reliably.

%
%
\begin{figure*}
  \centering
  \includegraphics[width=0.49\linewidth, angle=0]{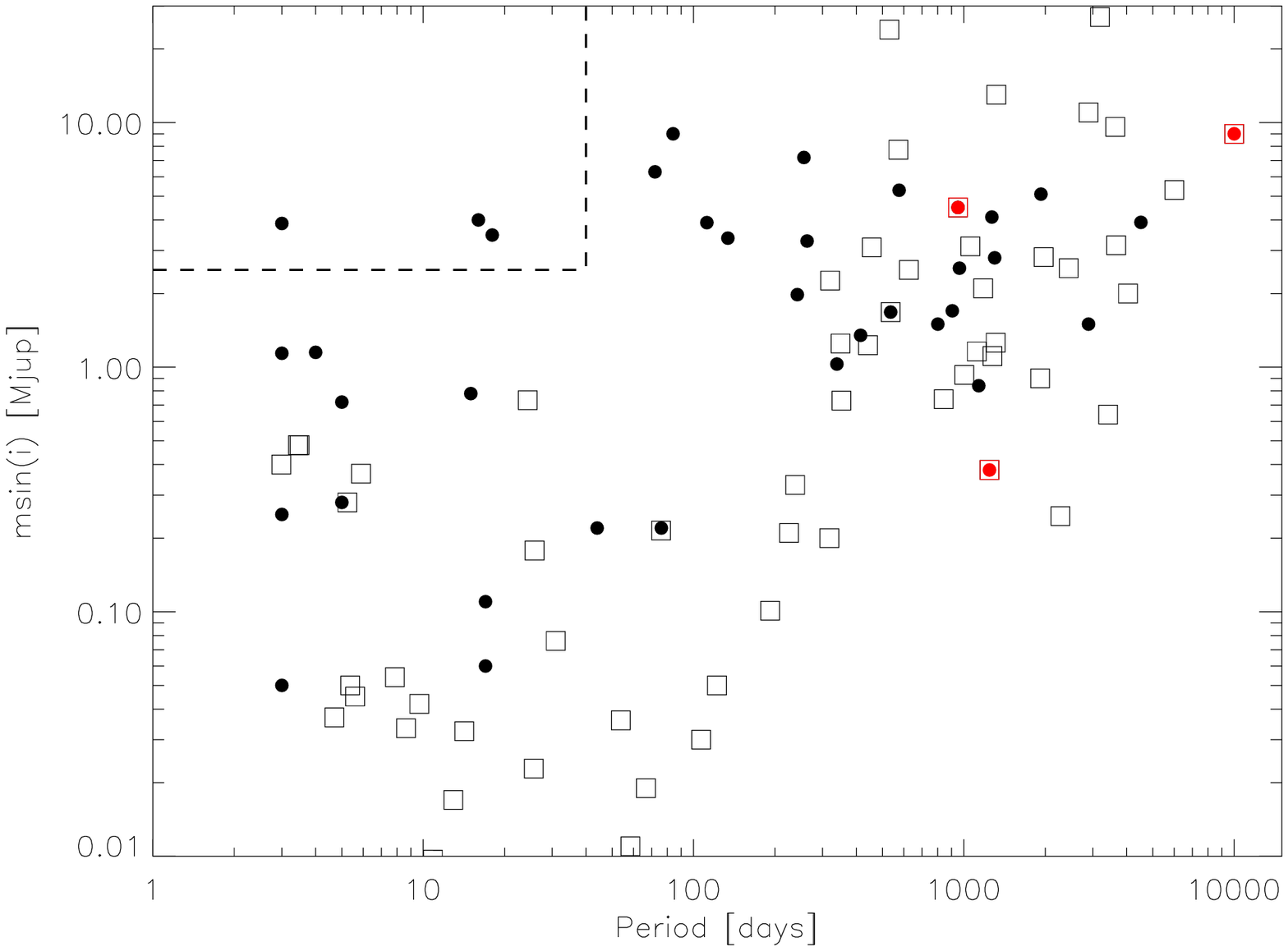}
  \includegraphics[width=0.49\linewidth, angle=0]{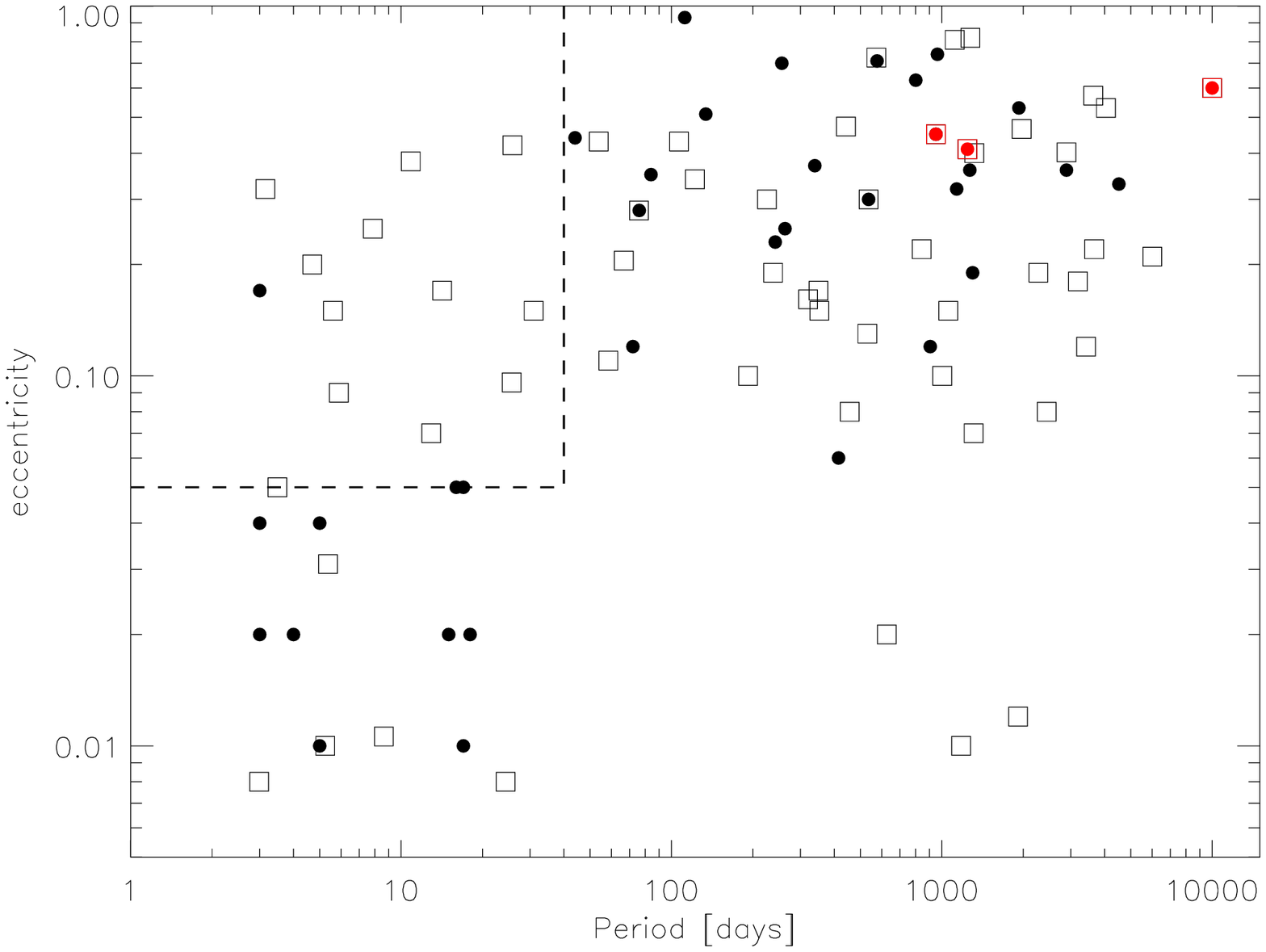}
  \includegraphics[width=0.49\linewidth, angle=0]{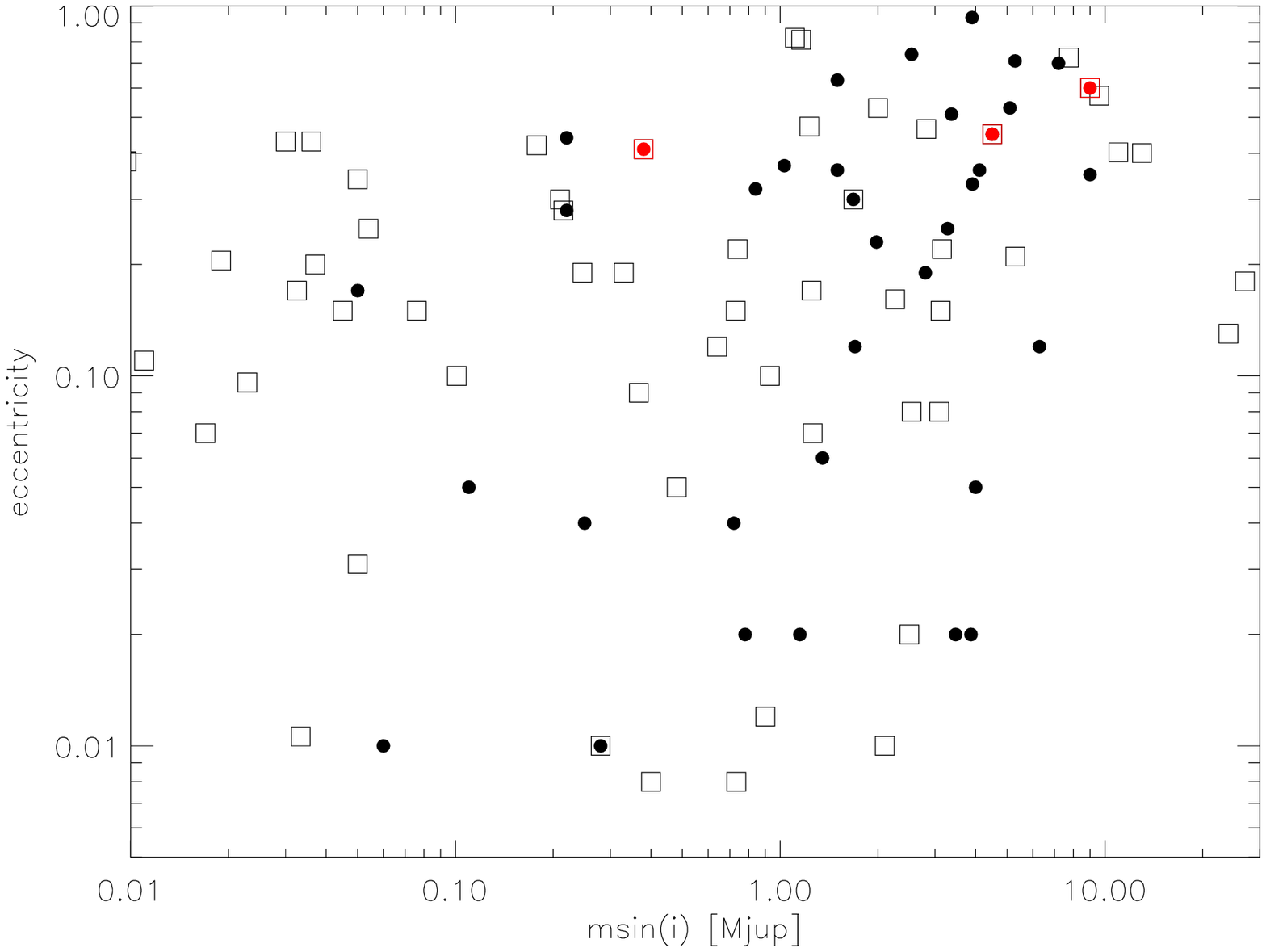}
  \includegraphics[width=0.49\linewidth, angle=0]{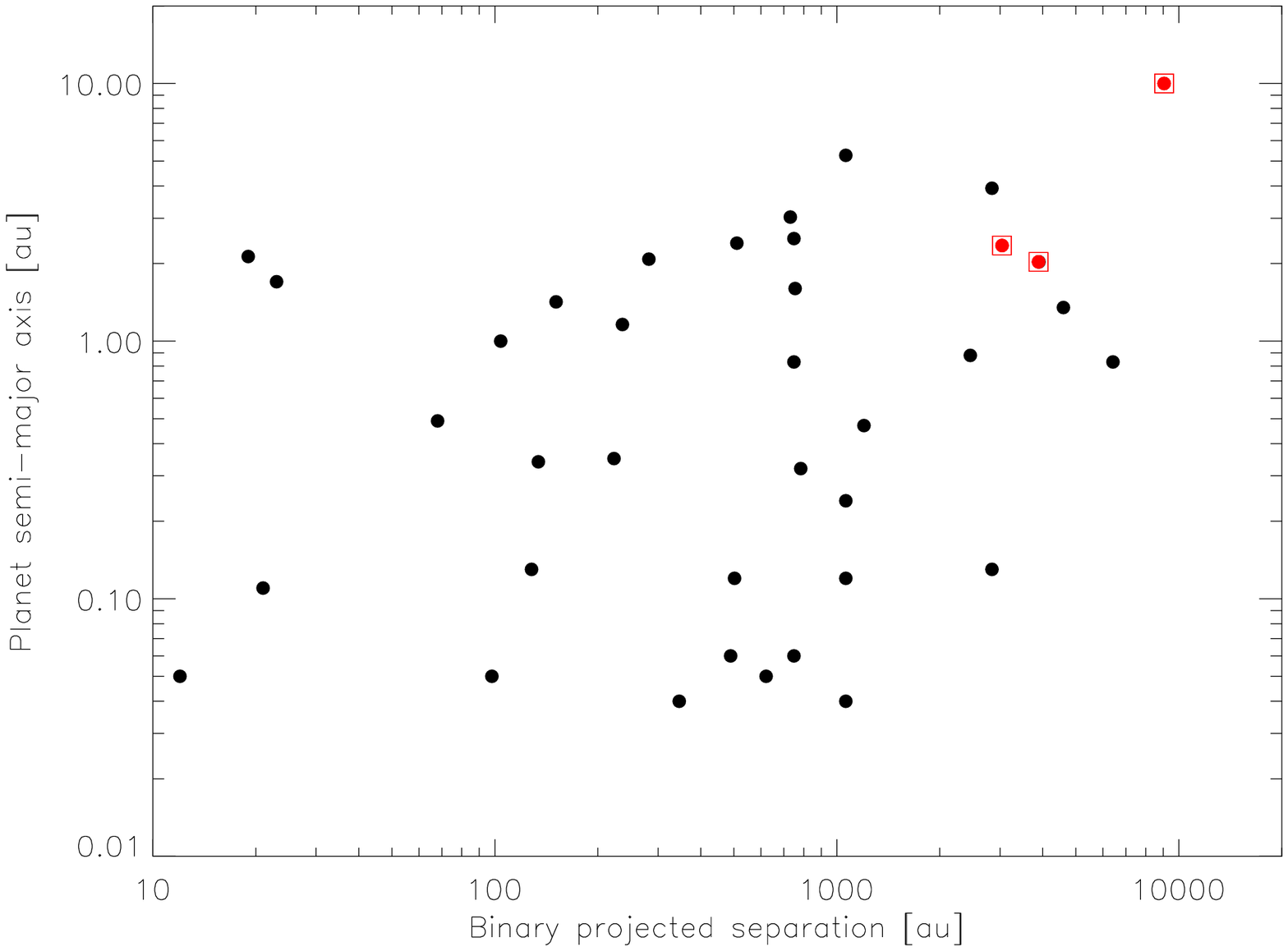}
   \caption{Diagrams showing the period-mass (top left), period-eccentricity
   (top right), mass-eccentricity (bottom left), and binary projected separation
   vs planet semi-major axis (bottom right) relations for single stars with
   planets within the VHS area considered in this work and multiple-star
   systems with planets (filled symbols). Black dots represent known
   planet-host stars previously published in the literature whereas
   red dots with open squares show our three best new discoveries.
   Plots amended from the original version published by \citet{eggenberger04}
   and later updated by \citet{mugrauer07a}, the lower right plot being added.
   }
   \label{fig_compHIP:Period_Mass_Eccentricity}
\end{figure*}
%

%
%
\section{Multiplicity of planet-host stars}
\label{compHIP:discussion_BF}

We have cross-correlated approximately 6200 square degrees of the Southern
sky imaged by VHS and 2MASS\@. We found 37 planet-host stars (20 G, 7 F, 10 K, 5 M)
with a full VHS covereage within a radius of 60 arcmin and separations larger than 2 arcsec 
between 2MASS and VHS\@. In this complete sample, we confirmed spectroscopically 
one substellar companion, two M dwarf companions, and another four (or five if
we add the companion to 91\,Aqr) at separations 
larger than 100,000 au. If we consider only dynamically-stable systems (see
Section \ref{compHIP:stability}), we derive a frequency of wide ($>$50--150 au)
stellar and substellar companions to planet-host stars of 2/37\,=\,5.4$\pm$3.8\% and 
1/37\,=\,2.7$\pm$2.7\% (1$\sigma$ confidence level with poissonian errors), respectively. 
Note that we have excluded from this estimate the known companions described in 
Section \ref{compHIP:surveys_completeness} because we do not recover them in our study 
(only two have projected separations larger than our lower limit).
In any case, our multiplicity estimates represent lower limits to the true 
binary frequency of planet-host stars because our study is not sensitive
to close systems \citep[e.g.\ HD\,126614 (K0$+$M;][]{howard00a}. 

Overall, our binary frequencies of a few percents are in agreement with
the number of wide substellar companions to young solar analogues 
\citep[0.5--6.3\% for separations in the 28--1590 au range;][]{metchev09a},
M dwarfs \citep[0.1--2.2\% for separations larger than $\sim$120 au ;][]{radigan08},
the frequency of 1$\pm$1\% of brown dwarf companions to GKM stars in the 
75--300 au range \citep{mccarthy04} or the lack of planetary-mass objects
more massive than 5--10 M$_{\rm Jup}$ around young nearby stars beyond
36--65 au \citep{masciadri05}. 
Similarly, \citet{raghavan10} derived a stellar multiplicity of 7.3\% and 2.2\%
among solar-type stars within 25 pc for projected separations in the 1000--10,000 au 
and 10,000--50,000 au ranges, respectively (their Figure 11 and Table 5).
These estimates are in good agreement with our results.
If we apply our photometric and astrometric selection constraints to the 454
solar-type stars listed in \citet{raghavan10}, we found 11 wide common proper motion
candidates to 74 stars (21 F, 33 G, 20 K) imaged by the VHS at the time of writing,
corresponding to 11/74\,=\,13.5$\pm$4.0\% if all are true companions.
Only one candidate lies close to its G-type primary whereas the others lie at projected 
physical separations larger than 10,000 au. This value is slightly larger 
than our determination, but in agreement within the error bars.
In conclusion, the frequency of wide stellar companions to solar-type stars
with or without planets (alt.\ close-in planets, radial velocity planets)
seems similar whereas there might be an excess of wide substellar companions
to planet-host stars, although this should be taken with caution because
this determination is based on one case only.

%
%
\section{Dynamical properties planet-host stars}
\label{compHIP:properties}

In Fig.\ \ref{fig_compHIP:Period_Mass_Eccentricity}, we plot three diagrams depicting
the period-mass, period-eccentricity, and mass-eccentricity relations for planet-host
stars in multiple systems (filled symbols) or isolated (open symbols). We have only
included the 49 planet-host stars (open symbols) in the area in common between VHS and 2MASS\@.
The planet-host stars with wide companions (filled black circles) are taken from 
\citet{eggenberger04} and \citet{mugrauer07a} and our new discoveries added as
red dots with open squares.
At the bottom right-hand side of Fig.\ \ref{fig_compHIP:Period_Mass_Eccentricity}
we have added a diagram showing the planet semi-major axis as a function of the
binary projected separation of known planet-host stars from \citet{mugrauer07a}
and our three new discoveries. We see a homogeneous distribution for the range
shown. We observe that our survey fills the top right
part of this diagram, extending the discovery to long periods and large binary
separations with a special mention to the substellar companion.

\citet{zucker02} and \citet{eggenberger04} found two main trends in the period-mass
diagram (top left panel in Fig.\ \ref{fig_compHIP:Period_Mass_Eccentricity}), facts
supported by the enhanced number of multiple systems discussed in \citet{mugrauer07a}.
On the one hand, no stars with planets more massive than 2.5 Jupiter masses
are found at periods less than 40 days. On the other hand, stars in multiple
systems have planets less massive than 5 M$_{\rm Jup}$ for periods longer than
100 days. Our discoveries confirm the latter trend where the systems with the
most massive planets exhibit periods longer than 950 days. We also found a
wide companion to HD\,126614, orbited by a small planet with a period longer than 
than 1000 days. However, the former
trend does not seem to hold because over 80 short period ($<$40 days) planets 
more massive than 2.5 M$_{\rm Jup}$ have now been announced around stars
by several programs like the COnvection ROtation and planetary Transits (Corot)
\citep{barge06a} and Kepler \citep{borucki10} missions or the 
Hungarian-made Automated Telescope \citep[HATNet;][]{bakos02} and
SuperWASP \citep{pollacco06} projects, among others.
These objects are not plotted in fig.\ \ref{fig_compHIP:Period_Mass_Eccentricity}
because they lie outside the VHS coverage but half of them should be observed 
by the VHS during the next years because they are in the Southern hemisphere.
We note that only four of these stars with short-period planets more massive
than 2.5 M$_{\rm Jup}$ were known prior to 2007, at the time of \citet{mugrauer07a}
publication. Moreover, most of these stars have small proper motions, typically
less than 0.2 arcsec/yr (our lower limit for the astrometric search), which
makes the search for co-moving companions extremely hard. This bias can
be easily understood in the case of CoroT, Kepler, HAT,Net and SuperWASP surveys,
which are dedicated to the search for planetary transits without a specific 
requirement in proper motion. 

We observe that the trend for low eccentricities and short periods among planet-host 
multiple systems holds (top right panel in Fig.\ \ref{fig_compHIP:Period_Mass_Eccentricity})
because we do not find any new candidate in this part of the diagram.
We note that only one system, not included in
the plots shown by \citet{mugrauer07a}, has a short period ($<$40 days)
and an eccentricity higher than 0.05, the possible planet "e" around 55\,Cnc\,A
\citep{wisdom05} which remains to be confirmed. We also find that our
discoveries confirm the trend seen in the mass-eccentricity diagram
(bottom left panel in Fig.\ \ref{fig_compHIP:Period_Mass_Eccentricity}), 
where more massive planets have on average higher mean eccentricities.
Overall, we corroborate the conclusions of \citet{desidera07} that
wide companions seem to have less influence on the properties of planets 
than closer systems. 

%
%
\section{Conclusions and future work}
\label{compHIP:conclusions}

We identified new potential common proper motion wide companions to known stars 
harbouring extrasolar planets. We confirmed spectroscopically two M dwarf and one 
mid-T dwarf companions sharing the same distance and proper motion as the planet-host star.
One companion to 91 Aqr remains as a candidate until radial velocity is
available. Moreover, four candidates lie further than 100,000 au suggesting
that these systems are not bound but might originate from a common stellar
origin, share a similar dynamical evolution, or are disrupted binaries drifting
away slowly from each other. The highlight of our search is 
the discovery of the second brown dwarf companion to a planet-host star, HIP\,70849, 
which might be an equal-mass binary. Overall, our success rate in confirming wide common
proper motion companions lies between 19\% and 50\%.
We derive frequencies of stellar and substellar wide companions to planet-host
stars of 5.4$\pm$3.8\% and 2.7$\pm$2.7\%, respectively.

Considering the success of our search for wide companions to planet-host stars
in a few thousands square degrees of the Southern sky imaged by the VHS, we plan 
to extend our automatic selection procedure to the entire Southern sky at the end 
of the VHS lifetime. Moreover, we plan to apply our method to Northern sources
cross-matching the UKIRT Infrared Deep Sky Survey \citep[UKIDSS;][]{lawrence07}
with 2MASS to provide a full view of the binarity among planet-host stars.
Additional studies of these new systems should include high-resolution imaging with
adaptive optics techniques to look for close companions and radial velocity
or transit searches to detect any additional exoplanet(s) in those systems.

%
%
\begin{acknowledgements}
NL was funded by the Ram\'on y Cajal fellowship number 08-303-01-02\@.
NL and VJSB are financially supported by the projects AYA2010-19136 and
AYA2010-20535 from the Spanish Ministry of Economy and Competitiveness 
(MINECO), respectively.
APG has been supported by Project No.\ 15345/PI/10 from the Fundaci\'on 
S\'eneca and MINECO under the grant AYA2011-29024\@.

This work is based on observations collected at the European Southern 
Observatory, Chile, under programme numbers 090.C-0791(A), 091.C-0862(B),
and 092.C-0874\@. Based on observation obtained as part of the VISTA Hemisphere 
Survey, ESO progamme, 179.A-2010 (PI: McMahon).

This work is based on observations (program GTC65-13B; PI Lodieu) made 
with the Gran Telescopio Canarias (GTC), operated on the island of La Palma 
in the Spanish Observatorio del Roque de los Muchachos of the Instituto de 
Astrof\'isica de Canarias.

The VISTA Data Flow System pipeline processing and science archive are
described in \cite{irwin04} and \citet{cross12}. We have used data from the
first data release (VHS DR1) described in \citet{mcmahon12}.
We are grateful to the Cambridge Astronomy Survey Unit and the VISTA 
Science Archive at the Wide Field Astronomy Unit, Edinburgh, for providing 
us with the reduced data and catalogues.

This research has made use of the Simbad and Vizier databases, operated
at the Centre de Donn\'ees Astronomiques de Strasbourg (CDS), and
of NASA's Astrophysics Data System Bibliographic Services (ADS).

This publication makes use of data products from the Two Micron
All Sky Survey (2MASS), which is a joint project of the University
of Massachusetts and the Infrared Processing and Analysis
Center/California Institute of Technology, funded by the National
Aeronautics and Space Administration and the National Science Foundation.

This research has benefitted from the exoplanet.eu webpage and the 
SpeX Prism Spectral Libraries, maintained by Adam Burgasser at 
http://pono.ucsd.edu/$\sim$adam/browndwarfs/spexprism.
\end{acknowledgements}
%

%
%
\bibliographystyle{aa}
\bibliography{../../AA/mnemonic,../../AA/biblio_old}
%

%
%
\appendix

\section{Potential moving group members}
\label{compHIP:new_comp_appendix}

In this appendix, we describe the properties of the planet-host stars
with very wide companions that might belong to moving group as well as 
systems rejected as bona-fide pairs based on our spectroscopic follow-up 
(Section \ref{compHIP:spectro}; Table \ref{tab_compHIP:list_candidates}).

\subsection{BD\,10$-$3166}
\label{compHIP:BD10-3166}

\citet{butler00} announced a planet with almost circular orbit (e\,=\,0.05) 
and a mass of 0.48 Jupiter masses
(M$_{\rm Jup}$) with a period of $\sim$3.5 days around the G4V solar-type star 
BD\,10$-$3166 \citep{hog98,hog00}. The star is located at 66 pc from the Sun,
has an age of $\sim$4.2 Gyr, is metal-rich ([Fe/H]\,=\,0.5 dex), and has a 
proper motion of $\sim$0.2 arcsec/yr. We confirm this
estimated distance based on the absolute magnitude of a mid-G dwarf
\citep[e.g.\ Figure 10 of][]{kirkpatrick12}.

We found a bright and wide common proper motion companion located at 
$\sim$39 arcmin from BD\,10$-$3166\@. We classify this potential companion
as a M5--M5.5 dwarf based on the GTC optical spectrum. 
Using the $J$-band magnitudes of M5 and M5.5 dwarfs with their errors
listed in Table \ref{tab_compHIP:dM_absMag} above, we derive a spectroscopic
mean distance of 75--77 pc with a possible range of 63--90 pc.
We conclude that these two objects are located at the same spectroscopic distance,
with a projected physical separation of $\sim$150,000--200,000 au.

\subsection{HD\,109749}
\label{compHIP:HD109749}

\citet{fischer06} discovered a planet with a mass of 0.28 M$_{\rm Jup}$
with a period of $\sim$5.2 days (e\,=\,0.01) around the G3IV (or G3V C)
star, HD\,109749 \citep{houk78}. The planet-host star is located at
56.3$\pm$4.3 pc with an age range of 7.4--13.2 Gyr and a metallicity estimated
to [Fe/H]\,=\,0.25$\pm$0.05 dex. The primary is part of a double system
formed by CCDM J12373-4049A and CCDM J12373-4049B, which are separated 
by $\sim$8.4--9.2 arcsec \citep{mason01}.

We detected two potential co-moving companions at 39.2 and 55.5 arcmin.
We classify the closest wide companion as a K7 dwarf, comparing its spectrum 
with HD\,157881 \citep{koen10}\footnote{http://www.ast.obs-mip.fr/users/leborgne/stelib/stars.html}.
Using Gl825 as a K7V reference at 3.946$\pm$0.012 pc 
\citep{hawley96,vanLeeuwen07}, we determine a spectroscopic distance 
of  approximately 140 pc. This distance estimate is 2.4 times greater than
the distance of the planet-host star, implying that these two objects do not form
a pair, even if we include the uncertainty on spectral classification and 
binary effects.

We classify the second companion as a M3.5$\pm$0.5 dwarf by direct comparison
with Sloan templates. We derive a mean spectroscopic distance of 73.3 pc for the
potential companion, with a range of 45--144 pc including the uncertainties on the
spectral type and absolute magnitude vs spectral type relations.
Hence, we conclude that these two objects are located at the same spectroscopic distance.

\subsection{HD\,143361}
\label{compHIP:HD143361}

\citet{jenkins09} detected an eccentric planet (e\,=\,0.17$\pm$0.15)
with a mass of 3.12 M$_{\rm Jup}$ and a orbital period larger than
1000 days around HD\,143361 \citep{houk78,jenkins11}. The planet-host
star is a G6V star \citep{houk78} at 65.66$\pm$5.51 pc from the Sun 
\citep{vanLeeuwen07} and a metallicity of Fe/H]\,=\,0.29 dex \citep{nordstroem04}.
We note that \citet{minniti09} quotes a spectral type of G0V (no reference is
given though) and a distance
of 59.35 pc for HD\,143361 based on the early analysis of Hipparcos parallax
\citep{perryman97}. Here we favour the new reduction of Hipparcos data
performed by \citet{vanLeeuwen07}.

We found two wide companions at 33.7 and 45 arcmin from the planet-host star.
We classify the closest candidate as a M6--M6.5 dwarf, suggesting a distance of
63--68 pc.
Both components seem to be located at a very similar spectroscopic distance so 
we conclude that they might be physically associated. At the distance of 33.7 arcmin, 
we derive a projected physical separation of $\sim$120,000--137,000 au for the pair.

We spotted a second potential co-moving companion at 45 arcmin from the planet-host
star. We discarded that object photometrically. Nonetheless, we took an optical
spectrum with NTT/EFOSC2 to confirm that it is an early-type star. Hence, this
candidate companion is not associated with the planet-host star.

\subsection{HD\,159868}
\label{compHIP:HD159868}

\citet{wittenmyer12} announced a second planet orbiting HD\,159868 
\citep{houk78} in addition to an eccentric planet candidate reported 
by \citet{Otoole07}. After collecting more data and re-analysing 
the periodiograms of HD\,159868, \citet{wittenmyer12} concluded that
this subgiant star at 58.7$\pm$2.7 pc harbours two planets with masses of
2.1 and 0.73 M$_{\rm Jup}$ and periods of about 1180 and 353 days,
respectively. The age of HD\,159868 is estimated to be 8.1 Gyr with
solar composition.

We uncovered two potential wide companion to the planet-host star,
at about 40.7 and 42.3 arcmin. We classify the closest as a M3.0$\pm$0.5
dwarf and the furthest as a sdM0.5 subdwarf. 
For the former, we derive a mean spectroscopic distance of 130 (72--175) pc,
twice larger than the distance of the planet-host star.

For the latter we infer a spectroscopic distance of 96$\pm$12 pc based
on the parallax of LHS\,536 \citep[44.6$\pm$5.3 pc;][]{vanAltena95}
and its spectral type \citep[sdM0.5;][]{gizis97a}. This distance is
much larger than the distance of the planet-host star. Moreover, 
we observe a metallicity difference between this candidate and the 
planet-host star.

To summarise, we reject both potential companions as co-moving objects
to the planet-host star HD\,159868\@.

\subsection{HD\,207832}
\label{compHIP:HD207832}

\citet{haghighipour12} discovered two jupiter mass planets orbiting
the solar-metallicity (Fe/H\,=\,0.06) G5V star HD\,207832 
\citep[54.4$\pm$2.7 pc;][]{vanLeeuwen07,jenkins11} 
by radial velocity. Their minimum masses are estimated to 0.56 and 
0.73 M$_{\rm Jup}$, with orbital periods of about 162 and 1156 days, 
and eccentricities of 0.13 and 0.27, respectively.

We unearthed a wide common proper motion companion to HD\,207832,
located at 38.6 arcmin. We classify it as a M6.5 dwarf, implying a distance of
48 pc (42--50 pc).

We conclude that these two objects most likely form a pair with similar spectroscopic
distances and a projected physical separation of $\sim$97,000--132,000 au.

\subsection{HD\,216435}
\label{compHIP:HD216435}

\citet{butler06} announced a new planet orbiting HD\,216435, a
G0V star at 32.62$\pm$0.42 pc \citep{vanLeeuwen07} with an 
age of 5.27 Gyr \citep{saffe05} and a metallicity of [Fe/H]\,=\,0.244\@. 
The planet, detected by radial velocity, has a minimum mass of 
1.26 M$_{\rm Jup}$, low eccentricity (e$\sim$0.07), and a period of
$\sim$1300 days.

We discovered a potential wide companion with $J$\,=\,12.631 mag at 
about 42 arcmin from HD\,216435\@. We classify it as a M4--M4.5 dwarf.
We derive a spectroscopic distance of 52--65 pc (41--82) for the companion,
suggesting that these two objects are not physically associated.

\subsection{91 Aqr}
\label{compHIP:91Aqr}

\citet{mitchell13} published a 3.2 M$_{\rm Jup}$ planet orbiting
the K0 giant, 91 Aqr. The planet-host star is located at 45.9$\pm$0.6 pc
\citep{vanLeeuwen07} with an age of 3.56$\pm$0.63 Gyr, and solar
composition. This planet-host star has the largest proper motion of 
our sample ($\sim$0.37"/yr). The planet has a low eccentricity (e\,=\,0.027) 
and a orbital period of 181.4 days.

We found a wide companion located at 52.9 arcmin from 91 Aqr and
a proper motion in good agreement with the planet-host star 
\citep{vanLeeuwen07}. We infer a spectroscopic distance of 63 (39--125) pc.
Hence, we can not discard the pair as wide co-moving
system, which would have a projected separation of $\sim$146,000 au. 
Radial velocity measurement are required to further assess the companionship.

\end{document}